\documentclass[twocolumn, trackchanges,floatfix]{aastex631}

\newcommand{\g}{$\gamma$-ray}

\definecolor{darkspringgreen}{rgb}{0.09, 0.59, 0.16}

\usepackage{subfigure}
\usepackage{amsmath,multirow, url,hyperref}
\usepackage[normalem]{ulem}
 \usepackage{booktabs}
\shorttitle{A new search for proton cascade emission in TXS 0506+056}
\shortauthors{The VERITAS Collaboration et al.}
\graphicspath{{./}{figures/}}
\begin{document}

\title{Probing a cosmogenic origin of astrophysical neutrinos and cosmic rays using gamma-ray observations of TXS 0506+056}
\shortauthors{The VERITAS Collaboration et al.}

\author[0000-0002-2028-9230]{A.~Acharyya}
\affiliation{CP3-Origins, University of Southern Denmark, Campusvej 55, 5230 Odense M, Denmark}
\author{A.~Archer}
\affiliation{Department of Physics and Astronomy, DePauw University, Greencastle, IN 46135-0037, USA}
\author[0000-0002-3886-3739]{P.~Bangale}
\affiliation{Department of Physics and Astronomy and the Bartol Research Institute, University of Delaware, Newark, DE 19716, USA}
\author[0000-0002-9675-7328]{J.~T.~Bartkoske}
\affiliation{Department of Physics and Astronomy, University of Utah, Salt Lake City, UT 84112, USA}
\author[0000-0003-2098-170X]{W.~Benbow}
\affiliation{Center for Astrophysics $|$ Harvard \& Smithsonian, Cambridge, MA 02138, USA}
\author[0000-0001-6391-9661]{J.~H.~Buckley}
\affiliation{Department of Physics, Washington University, St. Louis, MO 63130, USA}
\author[0009-0001-5719-936X]{Y.~Chen}
\affiliation{Department of Physics and Astronomy, University of California, Los Angeles, CA 90095, USA}
\author{J.~L.~Christiansen}
\affiliation{Physics Department, California Polytechnic State University, San Luis Obispo, CA 94307, USA}
\author[0000-0003-1716-4119]{A.~Duerr}
\affiliation{Department of Physics and Astronomy, University of Utah, Salt Lake City, UT 84112, USA}
\author[0000-0002-1853-863X]{M.~Errando}\affiliation{Department of Physics, Washington University, St. Louis, MO 63130, USA}
\author{M.~Escobar~Godoy}
\affiliation{Santa Cruz Institute for Particle Physics and Department of Physics, University of California, Santa Cruz, CA 95064, USA}
\author[0000-0002-5068-7344]{A.~Falcone}\affiliation{Department of Astronomy and Astrophysics, 525 Davey Lab, Pennsylvania State University, University Park, PA 16802, USA}
\author{S.~Feldman}
\affiliation{Department of Physics and Astronomy, University of California, Los Angeles, CA 90095, USA}
\author[0000-0001-6674-4238]{Q.~Feng}
\affiliation{Department of Physics and Astronomy, University of Utah, Salt Lake City, UT 84112, USA}
\author[0000-0002-2636-4756]{S.~Filbert}
\affiliation{Department of Physics and Astronomy, University of Utah, Salt Lake City, UT 84112, USA}
\author[0000-0002-1067-8558]{L.~Fortson}
\affiliation{School of Physics and Astronomy, University of Minnesota, Minneapolis, MN 55455, USA}
\author[0000-0003-1614-1273]{A.~Furniss}\affiliation{Santa Cruz Institute for Particle Physics and Department of Physics, University of California, Santa Cruz, CA 95064, USA}
\author[0000-0002-0109-4737]{W.~Hanlon}
\affiliation{Center for Astrophysics $|$ Harvard \& Smithsonian, Cambridge, MA 02138, USA}
\author[0000-0003-3878-1677]{O.~Hervet}
\affiliation{Santa Cruz Institute for Particle Physics and Department of Physics, University of California, Santa Cruz, CA 95064, USA}
\author[0000-0001-6951-2299]{C.~E.~Hinrichs}
\affiliation{Center for Astrophysics $|$ Harvard \& Smithsonian, Cambridge, MA 02138, USA and Department of Physics and Astronomy, Dartmouth College, 6127 Wilder Laboratory, Hanover, NH 03755 USA}
\author[0000-0002-6833-0474]{J.~Holder}\affiliation{Department of Physics and Astronomy and the Bartol Research Institute, University of Delaware, Newark, DE 19716, USA}
\author{Z.~Hughes}\affiliation{Department of Physics, Washington University, St. Louis, MO 63130, USA}
\author{M.~Iskakova}\affiliation{Department of Physics, Washington University, St. Louis, MO 63130, USA}
\author[0000-0002-1089-1754]{W.~Jin}
\affiliation{Department of Physics and Astronomy, University of California, Los Angeles, CA 90095, USA}
\author[0000-0002-3638-0637]{P.~Kaaret}
\affiliation{Department of Physics and Astronomy, University of Iowa, Van Allen Hall, Iowa City, IA 52242, USA}
\author{M.~Kertzman}
\affiliation{Department of Physics and Astronomy, DePauw University, Greencastle, IN 46135-0037, USA}
\author{M.~Kherlakian}
\affiliation{Fakult\"at f\"ur Physik \& Astronomie, Ruhr-Universit\"at Bochum, D-44780 Bochum, Germany}
\author[0000-0003-4785-0101]{D.~Kieda}
\affiliation{Department of Physics and Astronomy, University of Utah, Salt Lake City, UT 84112, USA}
\author[0000-0002-4260-9186]{T.~K.~Kleiner}
\affiliation{DESY, Platanenallee 6, 15738 Zeuthen, Germany}
\author[0000-0002-4289-7106]{N.~Korzoun}
\affiliation{Department of Physics and Astronomy and the Bartol Research Institute, University of Delaware, Newark, DE 19716, USA}
\author[0000-0003-4641-4201]{M.~J.~Lang}
\affiliation{School of Natural Sciences, University of Galway, University Road, Galway, H91 TK33, Ireland}
\author[0000-0003-3802-1619]{M.~Lundy}
\affiliation{Physics Department, McGill University, Montreal, QC H3A 2T8, Canada}
\author[0000-0001-9868-4700]{G.~Maier}
\affiliation{DESY, Platanenallee 6, 15738 Zeuthen, Germany}
\author[0000-0001-5937-446X]{C.~L.~Mooney}
\affiliation{Department of Physics and Astronomy and the Bartol Research Institute, University of Delaware, Newark, DE 19716, USA}
\author[0000-0002-3223-0754]{R.~Mukherjee}
\affiliation{Department of Physics and Astronomy, Barnard College, Columbia University, NY 10027, USA}
\author[0000-0002-6121-3443]{W.~Ning}
\affiliation{Department of Physics and Astronomy, University of California, Los Angeles, CA 90095, USA}
\author[0000-0002-4837-5253]{R.~A.~Ong}
\affiliation{Department of Physics and Astronomy, University of California, Los Angeles, CA 90095, USA}
\author[0000-0003-3820-0887]{A.~Pandey}
\affiliation{Department of Physics and Astronomy, University of Utah, Salt Lake City, UT 84112, USA}
\author[0000-0001-7861-1707]{M.~Pohl}
\affiliation{Institute of Physics and Astronomy, University of Potsdam, 14476 Potsdam-Golm, Germany and DESY, Platanenallee 6, 15738 Zeuthen, Germany}
\author[0000-0002-0529-1973]{E.~Pueschel}\affiliation{Fakult\"at f\"ur Physik \& Astronomie, Ruhr-Universit\"at Bochum, D-44780 Bochum, Germany}
\author[0000-0002-4855-2694]{J.~Quinn}
\affiliation{School of Physics, University College Dublin, Belfield, Dublin 4, Ireland}
\author{P.~L.~Rabinowitz}
\affiliation{Department of Physics, Washington University, St. Louis, MO 63130, USA}
\author[0000-0002-5351-3323]{K.~Ragan}
\affiliation{Physics Department, McGill University, Montreal, QC H3A 2T8, Canada}
\author{P.~T.~Reynolds}
\affiliation{Department of Physical Sciences, Munster Technological University, Bishopstown, Cork, T12 P928, Ireland}
\author[0000-0002-7523-7366]{D.~Ribeiro}
\affiliation{School of Physics and Astronomy, University of Minnesota, Minneapolis, MN 55455, USA}
\author{E.~Roache}
\affiliation{Center for Astrophysics $|$ Harvard \& Smithsonian, Cambridge, MA 02138, USA}
\author[0000-0003-1387-8915]{I.~Sadeh}
\affiliation{DESY, Platanenallee 6, 15738 Zeuthen, Germany}
\author{A.~C.~Sadun}
\affiliation{Department of Physics, University of Colorado Denver, Campus Box 157, P.O. Box 173364, Denver CO 80217, USA}
\author[0000-0002-3171-5039]{L.~Saha}
\affiliation{Center for Astrophysics $|$ Harvard \& Smithsonian, Cambridge, MA 02138, USA}
\author{G.~H.~Sembroski}
\affiliation{Department of Physics and Astronomy, Purdue University, West Lafayette, IN 47907, USA}
\author[0000-0002-9856-989X]{R.~Shang}
\affiliation{Department of Physics and Astronomy, Barnard College, Columbia University, NY 10027, USA}
\author[0000-0003-3407-9936]{M.~Splettstoesser}
\affiliation{Santa Cruz Institute for Particle Physics and Department of Physics, University of California, Santa Cruz, CA 95064, USA}
\author[0000-0002-9852-2469]{D.~Tak}
\affiliation{SNU Astronomy Research Center, Seoul National University, Seoul 08826, Republic of Korea.}
\author{A.~K.~Talluri}\affiliation{School of Physics and Astronomy, University of Minnesota, Minneapolis, MN 55455, USA}
\author{J.~V.~Tucci}
\affiliation{Department of Physics, Indiana University Indianapolis, Indianapolis, Indiana 46202, USA}
\author[0000-0002-8090-6528]{J.~Valverde}
\affiliation{Department of Physics, University of Maryland, Baltimore County, Baltimore MD 21250, USA and NASA GSFC, Greenbelt, MD 20771, USA}
\author[0000-0003-2740-9714]{D.~A.~Williams}
\affiliation{Santa Cruz Institute for Particle Physics and Department of Physics, University of California, Santa Cruz, CA 95064, USA}
\author[0000-0002-2730-2733]{S.~L.~Wong}
\affiliation{Physics Department, McGill University, Montreal, QC H3A 2T8, Canada}
\author{T.~Yoshikoshi}\affiliation{Institute for Cosmic Ray Research, University of Tokyo, 5-1-5, Kashiwa-no-ha, Kashiwa, Chiba 277-8582, Japan}
\collaboration{59}{The VERITAS Collaboration}

\author[0000-0002-0738-7581]{M.~Meyer}\affiliation{CP3-Origins, University of Southern Denmark, Campusvej 55, 5230 Odense M, Denmark}
\author[0009-0002-8425-9636]{J.~Müller}
\affiliation{CP3-Origins, University of Southern Denmark, Campusvej 55, 5230 Odense M, Denmark}
\affiliation{Department of Physics and Institute for Theoretical and Computational Physics, University of Crete, GR-70013 Heraklio, Greece }
\affiliation{Institute of Astrophysics, Foundation for Research and Technology– Hellas, Vassilika Vouton, GR-70013 Heraklio, Greece}
\collaboration{2}{The \emph{Fermi}--LAT Collaboration}

\correspondingauthor{Atreya Acharyya}
\email{atreya@cp3.sdu.dk}

\correspondingauthor{Connor Mooney}
\email{comooney@udel.edu}

\correspondingauthor{Manuel Meyer}
\email{mey@sdu.dk}
\begin{abstract}
In September 2017, a high-energy neutrino event detected by the IceCube Neutrino Observatory (IceCube-170922A) was associated, at the $3\sigma$ level, with a gamma-ray flare from the blazar TXS~0506+056. 
Cosmic rays that are accelerated in astrophysical sources can escape from their jets and interact with background radiation fields. Interactions with the extragalactic background light can produce pions and hence neutrinos, while interactions with the cosmic microwave background predominantly drive inverse Compton scattering, contributing to electromagnetic cascades in intergalactic space. The resulting secondary gamma-ray emission can be detected with high-energy gamma-ray telescopes. Here, we report on a new search for such cosmogenic cascade emission from the blazar TXS~0506+056, using a combined data set from the \textit{Fermi}--Large Area Telescope and VERITAS. We compare the gamma-ray spectrum and neutrino observations with the predictions of cosmic-ray induced cascades in intergalactic space. The observed gamma-ray spectrum is modeled as a combination of the primary spectrum and the cascade spectrum. We apply a Monte Carlo simulation with a $\Delta\chi^2$-based likelihood analysis to jointly determine the best-fit parameters of a proton emission spectrum describing the data and derive constraints on the proton escape luminosity. Assuming a log-parabola primary photon spectrum, we find consistency with a proton injection spectral index of $\alpha_{p} \simeq 2.0$ and a cutoff energy of $E_{p,\text{max}} \simeq 1.3 \times 10^{16}$ eV, and constrain the isotropic proton escape luminosity to $1 \times 10^{44}$ erg~s$^{-1}$  $\lesssim L_{p, esc} \lesssim 3 \times 10^{45}$ erg~s$^{-1}$ at the 90\% confidence level.
\end{abstract}




\section{Introduction} \label{sec:intro}

The origin of the diffuse astrophysical neutrino flux detected by the IceCube Neutrino Observatory at energies above 10 TeV remains an open question \citep{PhysRevLett.111.021103, PhysRevLett.113.101101}. 
High-energy neutrinos are expected to be produced through the interactions of cosmic-ray protons with ambient gas or radiation fields, leading to hadronic cascades. These same interactions can also generate high-energy gamma rays via the decay of neutral pions, providing a potential link between neutrino and gamma-ray emission. Such gamma rays are detectable with the \textit{Fermi}-Large Area Telescope (LAT; \cite{Fermi_LAT}) and ground-based imaging atmospheric Cherenkov telescopes (IACTs). In contrast, gamma-ray emission can also arise from inverse Compton (IC) scattering in leptonic scenarios, which do not produce neutrinos, emphasizing the need for multi-messenger observations to distinguish between models.


Several astrophysical source classes have been proposed and investigated as possible counterparts to the high-energy neutrinos. This includes blazars, a subset of active galactic nuclei (AGN) having relativistic particle jets closely aligned to our line-of-sight \citep{1989A&A...221..211M, 1991PhRvL..66.2697S, 1992PhRvL..69.2885P}.
Although the isotropic distribution of the astrophysical neutrino flux detected by IceCube suggests an extragalactic origin, the exact source remains uncertain. A study by \cite{2018AdSpR..62.2902A} found that the fractional contribution from \textit{Fermi}--LAT--detected blazars is less than 27$\%$. Other potential sources of astrophysical neutrinos include starburst galaxies, the Galactic plane, supernovae, clusters of galaxies, and tidal disruption events (for example \cite{Kurahashi2022, 2023Sci...380.1338I}).


The spectral energy distribution (SED) of a typical blazar comprises two broad emission features. The first peak, in the radio up to X-ray waveband, is commonly attributed to synchrotron emission from relativistic electrons and positrons in the jet. On the other hand, the origin of the second feature, in the X-ray to gamma-ray band, is less clear. Leptonic models \citep{Blandford_and_Levinson_1995, Georganopoulos_2002} attribute the second peak emission to the IC scattering between the energetic leptons in the jet and a field of low energy photons, either from the same synchrotron photons (synchrotron self-Compton [SSC] model) or from an external source (external inverse Compton [EIC] model). Hadronic models, conversely, suggest that the high-energy photons are produced in cosmic-ray interactions, primarily through the decay of neutral pions \citep{1992_Mannheim}, while proton synchrotron emission may provide a subdominant contribution \citep{2002_Aharonian}. The hadronic origin of the gamma-ray emission would potentially make AGN prime candidate sources of astrophysical neutrinos \citep{1995Mannheim, 1997Halzen}. 

Since 2016, IceCube has been broadcasting automatic real-time alerts of single high-energy ($>$ 60 TeV) neutrino events of potential astrophysical origin in order to allow for prompt follow-up observations. The information about these events is shared using the Gamma-ray Coordinates Network (GCN),\footnote{\url{https://gcn.gsfc.nasa.gov} (accessed on 05/23/2025)} an open-source platform created by NASA to receive and transmit alerts about astronomical transient phenomena within a minute after detection. These single-event alerts have a typical localization uncertainty of $\sim 1^{\circ}$ (68\% containment), and the region of interest (RoI), defined by this uncertainty, often contains potential neutrino sources such as AGN or transient sources.

On September 22, 2017, one such alert, IceCube-170922A, reported the detection of a high-energy neutrino with a most probable energy of $\sim$290~TeV \citep{GCN21916}. The event was classified as a \texttt{GOLD} alert, indicating a high probability of being astrophysical in origin, supported by a high neutrino signalness value that pointed to a strong likelihood of it being an astrophysical neutrino event.
This event was found to have a best-fit reconstructed direction within 0.1$^{\circ}$ of the sky position of a known blazar, TXS 0506+056, having a subsequently-measured redshift $z~=~0.337 \pm 0.001$ \citep{redshift_citation}. 
The correlation between the neutrino event and the gamma-ray flare of TXS~0506+056, measured with the \textit{Fermi}--LAT, was subsequently found to have a $> 3 \sigma$ statistical significance \citep{eaat1378}. This alert also prompted follow-up observations with ground-based instruments, leading to the first detection of this blazar in the very high-energy (VHE) regime with the MAGIC telescopes at a significance level of 6.2$\sigma$, based on 13 hours of observations between September 24, 2017 and October 4, 2017 \citep{ansoldi2018blazar}. 

TXS~0506+056 represents the first association between a neutrino alert and a flaring blazar and can be interpreted as evidence that blazars may accelerate cosmic rays capable of producing both neutrinos and gamma rays. However, the association significance remains modest ($\sim$3$\sigma$), and the origin of the gamma-ray emission, particularly during the high-energy flare, remains consistent with both leptonic and hadronic scenarios.
Furthermore, IceCube also reported a $3.5 \sigma$ excess of high-energy neutrino events with respect to atmospheric backgrounds at the location of TXS 0506+056 between September 2014 and March 2015, independent of and prior to the 2017 flaring episode \citep{2018_ICEcube_archive, 2018MNRAS.480..192P, 2019ApJ...880..103G}.
This was not accompanied by any detected gamma-ray flare, however.

In addition to the gamma rays and neutrinos produced at the source, there could be a contribution from the (so-called) cosmogenic neutrinos and gamma rays produced along the line-of-sight. 
In our analysis, these line-of-sight cascade photons are explicitly modeled using simulations and incorporated into the total spectrum fit. We restrict the contribution to photons arriving within a time delay consistent with the 2017 flare, and only include cascade photons arriving within the VERITAS source region. This procedure allows us to distinguish and constrain the cascade component jointly with the primary emission.
These gamma rays result from interactions between cosmic rays and background photons—specifically, photons from the extragalactic background light (EBL) and the cosmic microwave background (CMB). 

For example, the gamma rays produced from cosmic ray interactions ($p + \gamma \rightarrow \Delta^+ \rightarrow \pi^0 + p$; the $\pi^0$ decays immediately into two gamma rays) can interact with the photons of background radiation fields, $\gamma_{bg}$, to produce electron-positron pairs: $\gamma + \gamma_{bg} \rightarrow e^{+} + e^{-}$. 
These pairs can IC scatter the CMB photons up to gamma-ray energies and subsequently initiate electromagnetic cascades as the upscattered photons again pair-produce until the energy of the photons drop below the kinematic threshold for pair-production.  Gamma rays produced from electromagnetic cascades may have already been detected with IACTs, as photons generated through these proton-photon interactions can compensate for the high-energy TeV photons that are attenuated by EBL pair production, effectively populating the observed gamma-ray spectrum \citep{2010Essey, 2010APh....33...81E}.

In this paper, we report on a new search for cosmogenic proton cascade emission in TXS 0506+056, using a combined dataset from the \textit{Fermi}--LAT and the Very Energetic Radiation Imaging Telescope Array System (VERITAS). We compare the gamma-ray spectrum and neutrino observations with the predictions of cosmic-ray induced cascades in intergalactic space. 
We determine the best-fit parameters of a proton emission spectrum describing the data by modeling the observed gamma-ray spectrum as a combination of the primary spectrum, whose origin we remain agnostic about, and the cascade spectrum.
Our method has the advantage that the only assumption we make is that cosmic rays are accelerated in the source and subsequently escape. We also obtain constraints on the proton escape luminosity, using Monte Carlo simulations.

\section{Observations and Data Analysis}
\label{obs}
\subsection{VERITAS}
\label{sec:VERITAS_observations}

VERITAS is an array of four 12 m IACTs located at the Fred Lawrence Whipple Observatory (FLWO) in southern Arizona, USA ($31^{\circ}$ 40' N, $110^{\circ}$ 57' W, 1.3 km above sea level; \citealt{vts_paper}).  Each telescope is equipped with a camera containing 499 photomultiplier tubes and covering a field of view of 3.5$^{\circ}$. VERITAS is capable of detecting gamma rays having energies from 85 GeV to above 30 TeV, with an energy resolution of $ \Delta E/E~\sim$~15\% (at 1~TeV) and an angular resolution of $\sim$~0.1$^{\circ}$ (68\% containment at 1~TeV). 

VERITAS implements a wide-ranging neutrino follow-up program to search for potential VHE gamma-ray sources associated with IceCube neutrino events (for example \cite{santander2019recent, 2023_Qi, 2023arXiv230915469S, 2025ApJ...982...80A}). The follow-up program includes the long-term monitoring of TXS 0506+056, in response to the IceCube alert on September 23, 2017 (MJD 58019) and the reported flaring activity from the blazar with the \textit{Fermi}-LAT. The results from the initial VERITAS observations of TXS 0506+056, corresponding to a total exposure of 35 hours of quality-selected data, observed between September 23, 2017 and February 6, 2018, are reported in \cite{VERITAS_TXS0506_original}.

In this paper, we present an updated analysis of the initial follow-up, along with results from additional VERITAS observations of TXS 0506+056, including the long-term monitoring and an associated multiwavelength campaign. This includes a total of 114 hours of observations on TXS 0506+056 taken between MJD~57685 and MJD~60218, which corresponds to midnight on October~24, 2016 until midnight on October~1, 2023. 
Of this, 35 hours fall within the six-month flare period (September 23, 2017 – February 6, 2018) and are used in the cascade modeling. We define this period to ensure temporal consistency with the 2017 multi-messenger flare.
The VERITAS observations were performed using a standard ``wobble'' observation mode \citep{FOMIN1994151} with a $0.5^{\circ}$ offset in each of the four cardinal directions in order to allow for a simultaneous determination of the background event rate. Quality cuts were applied to the data set to remove observation periods affected by bad weather. The VERITAS data were analyzed using two independent analysis pipelines  \citep{Maier17, VEGAS}, yielding consistent results.

The VERITAS analysis parametrized the principal moments of the elliptical shower images and applied a set of cuts to these parameters to reject cosmic-ray background events. The cuts were determined using a boosted decision tree algorithm~\citep{2017APh....89....1K}, previously trained on gamma-ray shower simulations. They were optimized for soft-spectrum sources (spectral index $\sim 4$), as the expected VERITAS spectrum for TXS 0506+056 is steep above 100 GeV, consistent with previous VERITAS and MAGIC measurements \citep{VERITAS_TXS0506_original, ansoldi2018blazar}.

Only events with at least two telescope images are selected during this process. Gamma-ray candidate events are considered to fall in the source (\textsc{ON}) region if the squared angular distance, $\theta^{2}$, between the reconstructed event origin and TXS 0506+056 is less than 0.008 deg$^{2}$. The background was estimated using the ring background model \citep{2007A&A...466.1219B} where an annular ring surrounds the central source region.
An excess of $N_{\text{Excess}} = N_{\text{ON}} - \alpha N_{\text{OFF}} = $ 347 gamma-ray candidate events was recorded in the source region with $N_{\text{ON}}$ = 2136 \textsc{ON} events, $N_{\text{OFF}}$ = 16174 \textsc{OFF} events, and a background normalization factor $\alpha =$ 0.1106, corresponding to a statistical significance of 7.5$\sigma$, calculated following the method of \cite{1983ApJ...272..317L}. 
A lightcurve of the VERITAS flux above a threshold of 190~GeV and binned in monthly intervals is shown in Fig.~\ref{fig:lightcurve}. We note that the majority of the VERITAS significance ($\sim$8$\sigma$) was already achieved within the 35-hour flare interval, while the additional long-term monitoring increased the total exposure to 114 hours but only marginally raised the overall significance to 8.6$\sigma$.

\subsection{{\it Fermi}-LAT} 
\label{sec:Fermi_observations}

The \emph{Fermi}-LAT \citep{Fermi_LAT} is a pair conversion instrument capable of detecting gamma-ray photons in the energy range from 20~MeV to above 500 GeV. Primarily operating in survey mode, the \textit{Fermi}-LAT scans the entire sky every three hours. In this paper, we analyzed \textit{Fermi}-LAT data between MJD~54683 and MJD~60218. This corresponds to midnight on August 4, 2008, the start of the \textit{Fermi}-LAT mission, until midnight on October 1, 2023. Throughout the analysis, we use the \textit{Fermi} Science Tools version 2.2.0\footnote{\url{http://fermi.gsfc.nasa.gov/ssc/data/analysis/software} (accessed on 05/23/2025)} and \texttt{fermipy} version 1.2.2~\footnote{\url{http://fermipy.readthedocs.io} (accessed on 05/23/2025)} \citep{wood2017fermipy} in conjunction with the latest \textit{PASS} 8 IRFs~\citep{atwood2013pass}.

The \textit{Fermi}-LAT data were processed using a binned maximum likelihood analysis. Photons with energies between 100 MeV and 300 GeV that were detected within a RoI of radius 15$^{\circ}$ centered on the location of TXS 0506+056 were selected for the analysis. We selected only photon events from within a maximum zenith angle of 90$^{\circ}$ in order to reduce contamination from background photons from the Earth's limb, produced from the interaction of cosmic rays with the upper atmosphere.

All sources contained in the 4FGL-DR4 catalog \citep{2023_dr4} within a radius of $20^{\circ}$ from the source position of TXS 0506+056 were included in the model with their spectral parameters fixed to their catalog values. The contributions from the isotropic and Galactic diffuse backgrounds were modeled using the most recent templates for isotropic and Galactic diffuse emission, iso\_P8R3\_SOURCE\_V3\_v1.txt and gll\_iem\_v07.fits, respectively.  The normalization factors for both the isotropic and Galactic diffuse emission templates were left free along with the spectral normalization of all modeled sources within the RoI. Furthermore, the spectral shape parameters of all modeled sources within 3$^{\circ}$ of TXS 0506+056 were left free to vary while those of the remaining sources were fixed to the values reported in the 4FGL-DR4 catalog.  A spatial bin size of 0.1$^\circ$ per pixel and 4 energy bins per decade were used.

\begin{figure*}
\centering
\includegraphics[width=1\linewidth]{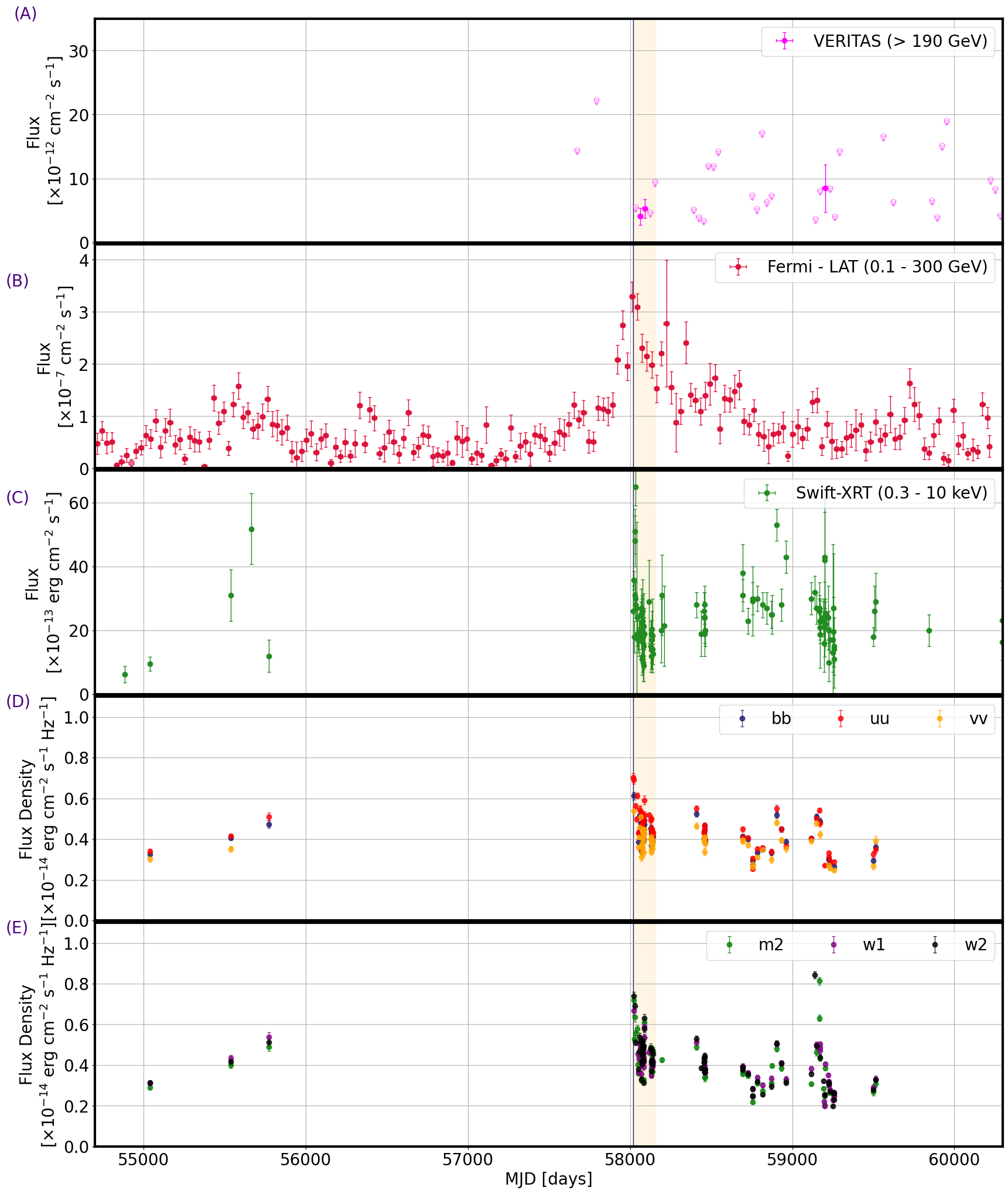}
\caption{Multiwavelength lightcurves of TXS 0506+056 between MJD~54683 and MJD~60218, which corresponds to midnight on August 4, 2008 until midnight on October 1, 2023. The blue narrow band corresponds to the IceCube alert, IceCube-170922A, observed on September 22, 2017 (MJD~58018). The orange shaded region corresponds to the time interval between September 23, 2017 and February 6, 2018, investigated in \cite{VERITAS_TXS0506_original}. \textbf{(A)} Monthly-binned VHE lightcurve for VERITAS observations above an energy threshold of 190 GeV. Upper limits at 95\% confidence level are quoted for observations with a significance of lower than $2\sigma$. \textbf{(B)} \emph{Fermi}-LAT lightcurve in 30 day bins in the 0.1--300 GeV band. \textbf{(C)} Swift-XRT energy flux in the 0.3–10 keV band. \textbf{(D and E)} Optical/UV flux observations with the \emph{Swift}-UVOT in the UW1, UW2, UM2, U, B, and V band filters, split into the UV and optical components for enhanced readability.
}
\label{fig:lightcurve}
\end{figure*}

TXS 0506+056 was detected at a high statistical significance of more than 129$\sigma$ (TS = 16738) during the observation period analyzed. 
The spectrum was found to be best modeled by a log parabola:
\begin{equation}
    \centering
   \frac{dN}{dE}=N_{0} \left(\frac{E}{E_0}\right)^{-\Gamma - \beta \text{ln} \left(\frac{E}{E_0}\right)}
	\label{LP_spectrum}
\end{equation}
where $N_{0}$ is the normalization, $E_{0}$ is the pivot energy, $\Gamma$ the spectral index and $\beta$ the curvature.
The best-fit spectral parameters obtained for TXS 0506+056 during the period investigated are $N_{0} = (6.59 \pm 0.11) \times 10^{-12}$ $\text{cm}^{-2}\text{s}^{-1} \text{MeV}^{-1}$, $\Gamma = 2.06 \pm 0.02$ and $\beta = 0.05 \pm 0.01$, with $E_{0} = 1126$ MeV. In order to investigate the temporal variability of the gamma-ray flux of TXS 0506+056 during the full time interval, the lightcurve of the integral flux in the energy range 100~MeV--300~GeV, shown in Fig.~\ref{fig:lightcurve}, is calculated with 30-day bins and keeping the spectral parameters fixed.

\subsection{{\it Swift}} 
\label{sec:Swift_observations}

The X-Ray Telescope (XRT; \citealt{XRT_Reduction}) is one of three co-aligned instruments aboard the \textit{Neil Gehrels Swift} Observatory. It is a grazing-incidence focusing X-ray telescope, capable of detecting  photons in the energy range between 0.2 and 10 keV \citep{2004AIPC..727..637G, 2005_Burrows}. TXS 0506+056 observations were performed with the \emph{Swift}-XRT between MJD~54683 and MJD~60218, which corresponds to midnight on August~4, 2008 until midnight on October~1, 2023.  A total observing time of 1370 ks was accumulated during this period in photon-counting (PC) mode and the data were analyzed using the \texttt{HEASOFT} v6.21 package. 

Each flux point is calculated using individual spectra fit with \texttt{XSPEC} over the energy range 0.3--10~keV. The spectra are fit using the Cash statistic, which is better suited than $\chi^2$ for low-count Poisson-distributed data, such as the case when dealing with a limited number of photons per spectral bin. A power-law model is employed for the fit, incorporating a Galactic absorption component fixed to the source direction with $N_{\mathrm{H}} = 1.11 \times 10^{21}$~cm$^{-2}$ \citep{column}.
An adequate fit for the combined spectrum ($C_{stat}/\mathrm{dof}$= 738/734) is found for a power law index of 1.99$~\pm$ 0.01 and normalization of $(5.57 \pm 0.06) \times 10^{-4}$ $\text{photons } \text{cm}^{-2}\text{s}^{-1} $. A lightcurve of the observed flux values is shown in Fig.~\ref{fig:lightcurve}.
The average flux, determined using a combined spectrum, is $(2.42 \pm 0.06) \times 10^{-4}$ erg cm$^{-2}$ s$^{-1}$. Photon pile-up at this rate level is negligible, and statistical uncertainties dominate over systematic ones. 

The Ultraviolet/Optical Telescope (UVOT), also onboard the \textit{Neil Gehrels Swift} Observatory, is a telescope sensitive to photons having energies ranging roughly between 1.9 eV and 7.3 eV \citep{2005SS_Roming}. \emph{Swift}-UVOT observations are performed in parallel to the \emph{Swift}-XRT measurements in six filters with central wavelengths of $V$ (5468 \AA), $B$ (4392 \AA), $U$ (3465 \AA), $UVW1$ (2600 \AA), $UVM2$ (2246 \AA), and $UVW2$ (1928 \AA). The \emph{Swift}-UVOT data were also analyzed with \texttt{HEASOFT} v6.21. A source region with a radius of $5.0''$ centered on TXS 0506+056, and a background region of the same size, away from the blazar and containing no known sources in any band, were used to extract signal and background counts, respectively. The magnitude of the source was then computed using \texttt{uvotsource} and converted to flux using the zero-point for each of the UVOT filters from \citet{Poole08}. Fig.~\ref{fig:lightcurve} also includes a lightcurve of the \emph{Swift}-UVOT observations.

\section{The Role of proton Cascade Emission}
\label{sec:cascade}

The association of the high-energy neutrino event IceCube-170922A detected with the IceCube Neutrino Observatory \citep{GCN21916} with a gamma-ray flare of TXS~0506+056 measured with the \textit{Fermi}--LAT \citep{eaat1378} suggests that blazars accelerate cosmic rays that produce neutrinos and gamma rays. As a result, TXS 0506+056 is a prime candidate to investigate proton cascade emission.
Accelerated cosmic rays can interact with matter and photon fields and produce gamma rays and neutrinos either at the source, in the intergalactic space, or both. It is possible that the cosmic rays escape while the observed gamma-ray emission is produced, for example via the SSC process of accelerated electrons. However, all cosmic rays that escape the source can potentially initiate particle cascades during the interaction with background radiation fields such as the CMB and the EBL (for example \cite{2013PhRvL.111d1103K}). Photon fields local to the blazar jet are not considered in this study, as we focus on the development of cascades in the intergalactic medium beyond the source.


In this paper, we self-consistently model the gamma ray and neutrino emission and the development of the full cosmic-ray cascade in the intergalactic medium using the \texttt{CRPropa} Monte Carlo simulation package version 3.1.6\footnote{\url{https://github.com/CRPropa/CRPropa3} (accessed on 05/23/2025)} \citep{2016JCAP...05..038A}. Initially, we assume that the cosmic rays accelerated by TXS 0506+056 are protons and we follow their propagation with the \texttt{CRPropa} code. We assume that these protons are injected into the intergalactic medium with a prescribed spectrum and directionality, escaping efficiently and without significant time delay along or near the line of sight. The protons produce neutrinos and gamma rays through photopion production and subsequent decay of the charged and neutral pions. Furthermore, they also create electron-positron pairs through Bethe-Heitler pair production. The produced gamma rays undergo ordinary, double, or triple pair production and the pairs produce further gamma rays through IC scattering, thus initiating the cascade. 

The simulations consider all relevant interactions and energy loss processes, namely pair production, IC scattering, synchrotron emission, and adiabatic losses due to the expansion of the Universe, and include the CMB, EBL and ultraviolet background radiation as target photon fields \citep{2020_Alves_Batista}. We consider only cascade photons that arrive within a time delay shorter than the duration of the blazar's active phase, \( t_{\text{max}} \). Following~\citet{VERITAS_TXS0506_original}, we define \( t_{\text{max}} = 6\) months as this includes the period from September 23, 2017 to February 6, 2018, during which the source was observed with VERITAS in a high state before transitioning to a low state in the \textit{Fermi}-LAT data. This time window is used throughout this work to constrain the proton cascade emission to be temporally consistent with the observed multi-messenger flare.

Using a spectral reweighting \citep{2018ApJS..237...32A} and parallel transport \citep{2016PhRvD..94h3005A}, the simulations are converted into energy-dependent sky maps for arbitrary input spectra and angles, $\theta_{\text{obs}}$, between the jet axis and the line of sight\footnote{A Python package performing these calculations is publicly available at \url{https://github.com/me-manu/simCRPropa} (accessed on 05/23/2025)} \citep{2023ApJ...950L..16A}. For simplicity, we assume $\theta_{\text{obs}} = 0$ throughout. The energy-dependent sky maps serve as input templates for the data analysis. 

The templates are generated for magnetic fields $B = 10^{-16}$, $10^{-15}$, and $10^{-14}$~G, reflecting plausible values of the intergalactic magnetic fields in cosmic voids, which are the dominant propagation environment over extragalactic distances \citep{2013A&ARv..21...62D}. For a fixed PSF aperture (0.1$^{\circ}$) and a 6-month delay window, the PSF-contained cascade flux decreases monotonically with $B$. We therefore highlight $B = 10^{-16}$~G as our benchmark, since in this case the electron--positron pairs undergo minimal deflection and the cascade remains contained within the PSF and ON region. Stronger fields broaden the cascade, reducing the contained flux and weakening the inferred constraints on the proton escape luminosity $L_{p}$. The \cite{Dominguez_EBL} model was used to model the EBL, with the redshift fixed at $z~=~0.337$ \citep{redshift_citation}. 

Stronger fields typical of galaxies or clusters are not included, as they are less relevant to the void-dominated paths considered in our cascade modeling. Moreover, for proton energies $\lesssim 10^{20}~\mathrm{eV}$, the mean free paths against interactions with the EBL and CMB are typically hundreds of megaparsecs or more \citep{2006PhRvD..74d3005B}, implying that local substructures such as galaxy clusters do not significantly alter their trajectories, and propagation remains dominated by cosmic voids.

For each intergalactic magnetic-field strength, we performed 1000 independent 
simulations and combined them into a single spectrum. A convergence study confirmed 
this choice: with $N=500$ trials the cascade spectrum exhibits residual fluctuations 
and underestimates the cascade contribution by nearly $75\%$, whereas increasing 
from $N=1000$ to $N=2000$ alters the integrated flux by only $\sim 6\%$. We therefore adopt $N=1000$ realizations per magnetic-field value as a practical compromise that yields smooth, reproducible spectra while keeping the runtime tractable ($\sim$3 days per run on the computing cluster).

An example of a simulated skymap is given in Fig.~\ref{fig:lp_sky}, where we show the simulated photon cascade flux per solid angle. The value assumed for the magnetic field is $B = 10^{-16} \text{G}$. Also shown for comparison are the PSF of the \textit{Fermi}-LAT and VERITAS. In order to account for the fact that VERITAS will only see a fraction of the cascade photons, we only consider cascade photons arriving within the VERITAS ON region when investigating the cascade component of the spectrum. For the sake of readability, in the following discussions, we primarily consider the results obtained for a magnetic field strength $B = 10^{-16}$ G. A coherence length of 1 Mpc is assumed throughout. For the jet opening angle, we consider the canonical relationship between the Lorentz factor, $\Gamma$, and the jet opening angle, $\theta$, derived in \cite{2005AJ....130.1418J}, $\theta = \rho / \Gamma$, where $\rho$ is a constant. Using the value $\Gamma = \delta/2 =12.1$ \citep{2020ApJ...891..115P}, where $\delta$ is the Doppler factor of the jet, and the best-fit value of $\rho = 0.17$, we obtain the jet opening angle,~$\theta~=~0.8^{\circ}$.

To speed up the computation process, the number of injected particles is reduced for larger injected energies. Moreover, all the secondary particles produced are tracked, with the thinning parameter set at $\eta = 0$ \citep{2016JCAP...05..038A}. To optimize the computational efficiency, particle tracing is terminated when their energy falls below 0.1 GeV or when their total propagation distance exceeds 4 Gpc. Additionally, we apply a minimum rigidity cut corresponding to 50 GV at injection, as electron-positron pairs below this produce IC scattered photons below the \textit{Fermi}--LAT energy band.
A standard $\Lambda$CDM cosmology is assumed with the default \texttt{CRPropa}  values, $H_{0} = 67.3$ km s$^{-1}$ Mpc$^{-1}$ and a matter density of $\Omega_{m} = 0.315$. 
We inject protons at discrete energies, using delta functions, in the energy range between $10^{12}$ eV and $10^{18}$ eV and simulate arbitrary injection spectra by re-weighting the number of particles in each energy bin with the flux of the desired spectrum. We choose a proton injection spectrum that follows a power law with a high energy exponential cut-off: 
\begin{equation}
\centering
\frac{dN_{\text{inj}}}{dE_{p}} = \kappa {E_{p}}^{-\alpha_{p}} \exp\left(- \frac{E_{p}}{E_{\text{p, max}}}\right), 
\label{casc_spectrum}
\end{equation}
where the “inj” and “p” subscripts represents the proton injection spectrum and the proton energy, respectively. $\kappa$ is a normalization factor kept free during the fitting procedure. 

The proton injection spectral slope parameter considered is in the interval $\alpha_{p}~\in~[1.8, 2.6]$ and the high energy cut-off interval is $E_{\text{p, max}}~\in~[10^{15}, 10^{19}]$~eV.  The higher end of the cut-off energy range is within the domain of UHECRs and even higher energies may be possible for nuclei heavier than protons. 
We were unable to simulate to higher cut-off energies due to the extensive computational resources required for those simulations.
Nevertheless we find, later, that the total $\chi^{2}$ statistic tends to increase with increasing $E_{\text{p,max}}$ beyond the region of best fit. This trend suggests that our conclusions about the optimal injection parameters remain robust even without simulations at significantly higher cut-off energies.

The combined spectrum is expressed as a sum of the primary spectrum and the cascade spectrum. The primary spectrum, $\frac{dN_{\text{prim}}}{dE_{\gamma}}$ (where the subscript represents energy of the gamma rays), is modeled first as a log parabola (LP):
\begin{equation}
    \centering
   \frac{dN_{\text{prim}}}{dE_{\gamma}}=N_{0} \left(\frac{E_{\gamma}}{E_0}\right)^{-\Gamma_{\text{LP}} - \beta \text{ln} \left(\frac{E_{\gamma}}{E_0}\right)} \text{exp}(-\tau_{\gamma\gamma}),
	\label{LP_spectrum2}
\end{equation}
where $N_{0}$ is the normalization, $E_{0}$ is the pivot energy, $\Gamma_{\text{LP}}$ is the spectral index and $\beta$ is the curvature.
The primary spectrum is also modeled as a power law with an exponential cut-off (PLE) of the form:
\begin{equation}
\centering
\frac{dN_{\text{prim}}}{dE_{\gamma}}=N_{0} \left(\frac{E_{\gamma}}{E_0}\right)^{-\Gamma_{\text{PLE}}}  \exp {\left ( - \frac{E_{\gamma}}{E_{\gamma, \text{cut}}}\right)} \text{exp}(-\tau_{\gamma\gamma}),
\label{prim_spectrum}
\end{equation}
where $E_{\gamma, \text{cut}}$ is the cut-off energy.
The high-energy cut-off reduces the number of gamma rays that can initiate a cascade. For both the primary spectrum models, we also include an EBL attenuation term, $\text{exp}(-\tau_{\gamma\gamma})$.

It should be noted that we remain agnostic about how the primary gamma rays are produced. While one could try to do everything self-consistently within the framework of a hadronic model, that is not our aim here. 
Furthermore, the contribution of proton synchrotron emission is negligible compared to the measured gamma-ray fluxes for our choice of magnetic fields (for example \cite{2022PhRvD.106l3005S}) and is not included in this study. Moreover, as seen in Fig.~\ref{fig:lightcurve}, TXS~0506+056 is a variable source and the VERITAS and \textit{Fermi}--LAT observations are not strictly contemporaneous.
The \textit{Fermi}-LAT and VERITAS spectral datasets are combined using \texttt{gammapy} \citep{gammapy:2023}. The combination was performed at the flux-points level, without a joint likelihood analysis. Systematic uncertainties were added in quadrature to the statistical errors, assuming 10\% for \textit{Fermi}-LAT and 20\% for VERITAS \citep{Aliu2014_1ES0229}, and these were propagated in the spectral modeling.


To determine the optimal proton injection parameters $(\alpha_{\text{p}}, E_{\text{p,max}})$ for a given magnetic field strength $B$, we perform a grid search over these parameters. For each pair $(\alpha_{\text{p}}, E_{\text{p,max}})$, we use \texttt{CRPropa} to simulate the resulting cascade gamma-ray spectrum produced by ultra-high-energy protons. This simulated cascade component is added to a parameterized primary gamma-ray spectrum.
We then fit this combined model to the observed \textit{Fermi}-LAT and VERITAS SED. In this fit, only the parameters of the primary spectrum and the normalization of the cascade component are varied, while $(\alpha_{\text{p}}, E_{\text{p,max}})$ are held fixed. The fit is performed using the \texttt{MINUIT} routine in \texttt{gammapy}, and the resulting $\chi^2$ is recorded for that grid point. By repeating this procedure across the full $(\alpha_{\text{p}}, E_{\text{p,max}})$ grid, we identify the optimal proton injection parameters as those corresponding to the minimum total $\chi^2$. This external grid search strategy allows us to systematically constrain the proton injection properties responsible for the observed gamma-ray spectrum.

The parameters kept free during the fitting are, therefore, the cascade component normalization $\kappa$, the normalization of the primary spectrum $N_{0}$, and either the spectral index and curvature ($\Gamma_{\text{LP}}, \beta$) for the LP model or the index and cut-off energy ($\Gamma_{\text{PLE}}, E_{\text{cut}}$) for the PLE model. 
The parameters describing the cosmic-ray spectrum, namely the spectral index $\alpha_{p}$ and the maximum proton energy $E_{p,\text{max}}$, were not treated as free parameters but instead held fixed during each fit. Their values were explored separately using a grid scan, with the optimal pair identified as the one yielding the minimum total $\chi^{2}$.

Two examples of the fit to the combined gamma-ray spectrum, one each for an LP and PLE primary spectrum respectively, are shown in Fig.~\ref{fig:lp_spectrum}. Also shown are the corresponding neutrino spectra and the representative neutrino flux upper limits that would result in, on average, one detection like IceCube--170922A over a period of 0.5 and 7.5 years, assuming an injection spectrum of $dN/dE \propto E^{-2}$ at the most probable neutrino energy of $E_{\nu} = 311$~TeV \citep{2018_Icecube}. It is worth noting, that earlier estimates placed the most probable energy at $E_{\nu}=290$ TeV and even as low as $E_{\nu}=120$ TeV in the GCN Circular \citep{GCN21916}.

\section{Results}

\subsection{Constraints on the spectrum of protons escaping the source}

\subsubsection{Log parabola primary photon spectrum}

\begin{figure}
\centering

\includegraphics[width= 1.1 \linewidth]{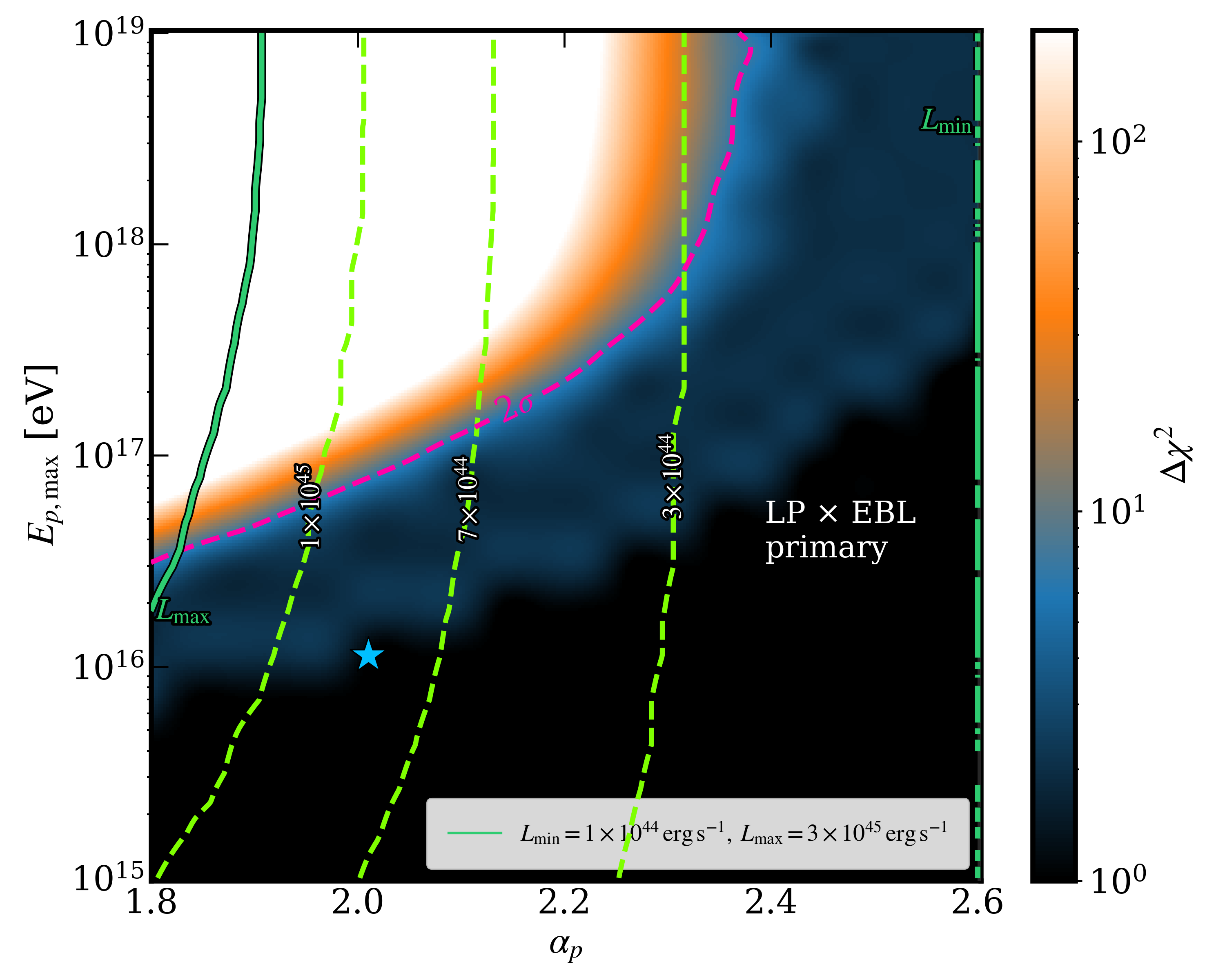}
\includegraphics[width= 1.1 \linewidth]{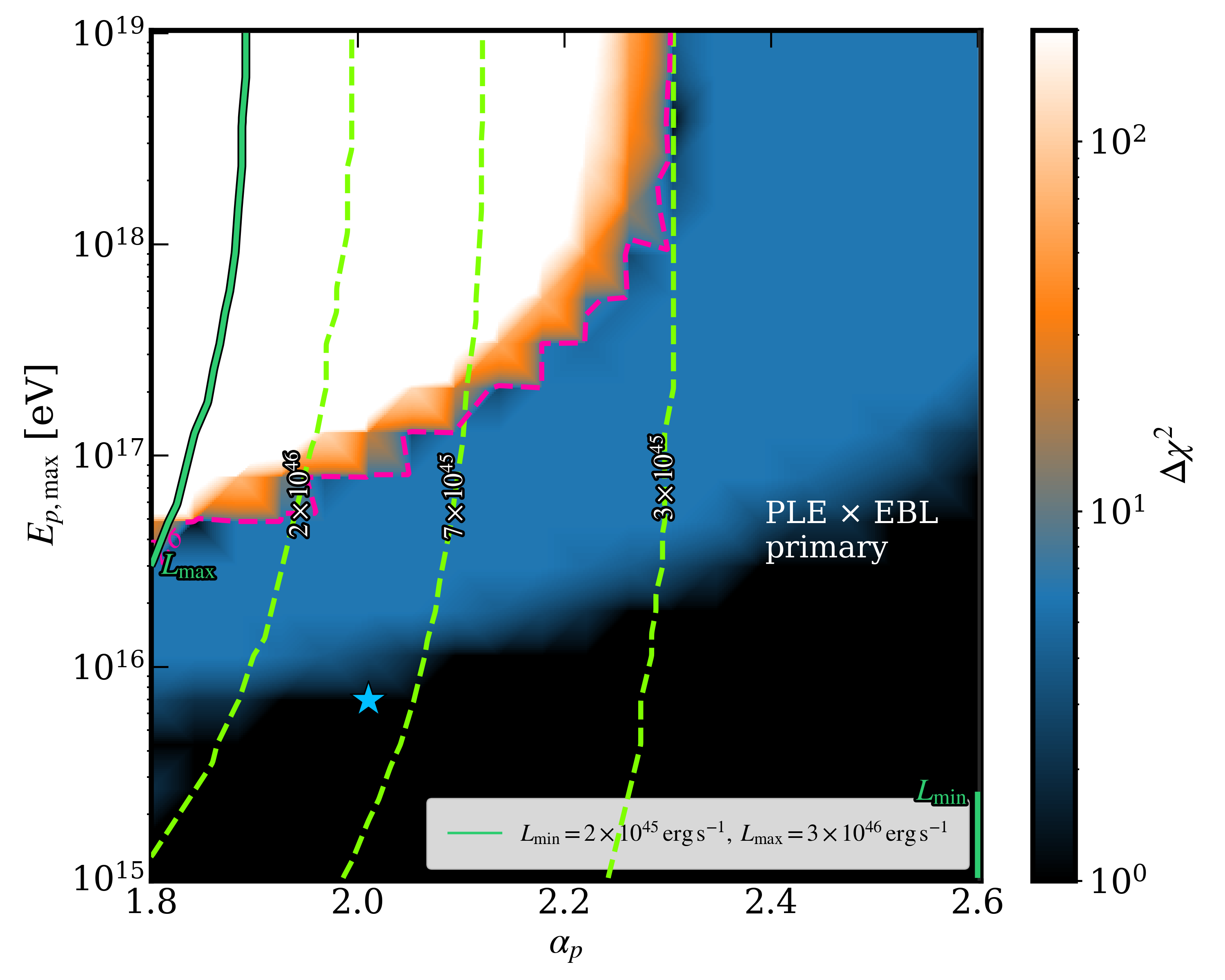}
\caption{The stats-surfaces obtained over the investigated parameter space of the proton spectrum, $E_{p, \text{max}}$ and $\alpha_{p}$, assuming a LP $\times$ EBL primary photon spectrum (top) and PLE $\times$ EBL primary photon spectrum (bottom). The values assumed for the maximum blazar activity time and magnetic field are $t_{\text{max}} = 6$ months and $B = 10^{-16} \text{G}$ respectively. The color scale denotes the $\Delta \chi^{2}$, the difference between the $\chi^{2}$ values at each point and the minimum $\chi^{2}$ value obtained over the entire parameter space. The blue stars mark the best-fit parameter combination corresponding to the global minimum $\chi^{2}$, and listed in Table~\ref{tab:table}. The dashed purple curves show 2$\sigma$ uncertainty contours. The dashed green curves represent points in the parameter space which correspond to the same proton escape luminosity, as labeled. The dark green curves mark the lower and upper bounds on the proton escape luminosity, denoted as $L_{\min}$ and $L_{\max}$, respectively and shown in the legend.}
\label{fig:lp_contour}
\end{figure}

On the top panel of Fig.~\ref{fig:lp_contour}, we show the $\Delta \chi^2$-surface obtained over the parameter space investigated for the proton injection parameters, $E_{\text{p, max}}$ and $\alpha_{p}$, assuming a LP primary gamma-ray spectrum with EBL absorption. The color scale denotes the difference between the $\chi^2$ value at each point and the minimum $\chi^2$ value obtained over the entire parameter space. Fig.~\ref{fig:lp_contour} also shows the 2$\sigma$ uncertainty contours, depicting regions of the parameter space not compatible with the gamma-ray data, based on the exclusion applied in this study. This includes, for example, a combination of a proton spectral slope $\alpha_{p} < 2$ and a high energy cut-off $E_{p, \text{max}} > 10^{17}$~eV. Assuming the Doppler factor of the acceleration site in the blazar jet is $\delta = 24.2$ \citep{2020ApJ...891..115P}, the compatible isotropic source-frame proton escape luminosities (corresponding to $\Delta \chi^2 \leq 2.71$ i.e. a 90$\%$ confidence level) are between $1 \times 10^{44}$ erg s$^{-1}$ and $3 \times 10^{45}$~erg~s$^{-1}$. 

\begin{figure}
\centering
\includegraphics[width= 1 \linewidth]{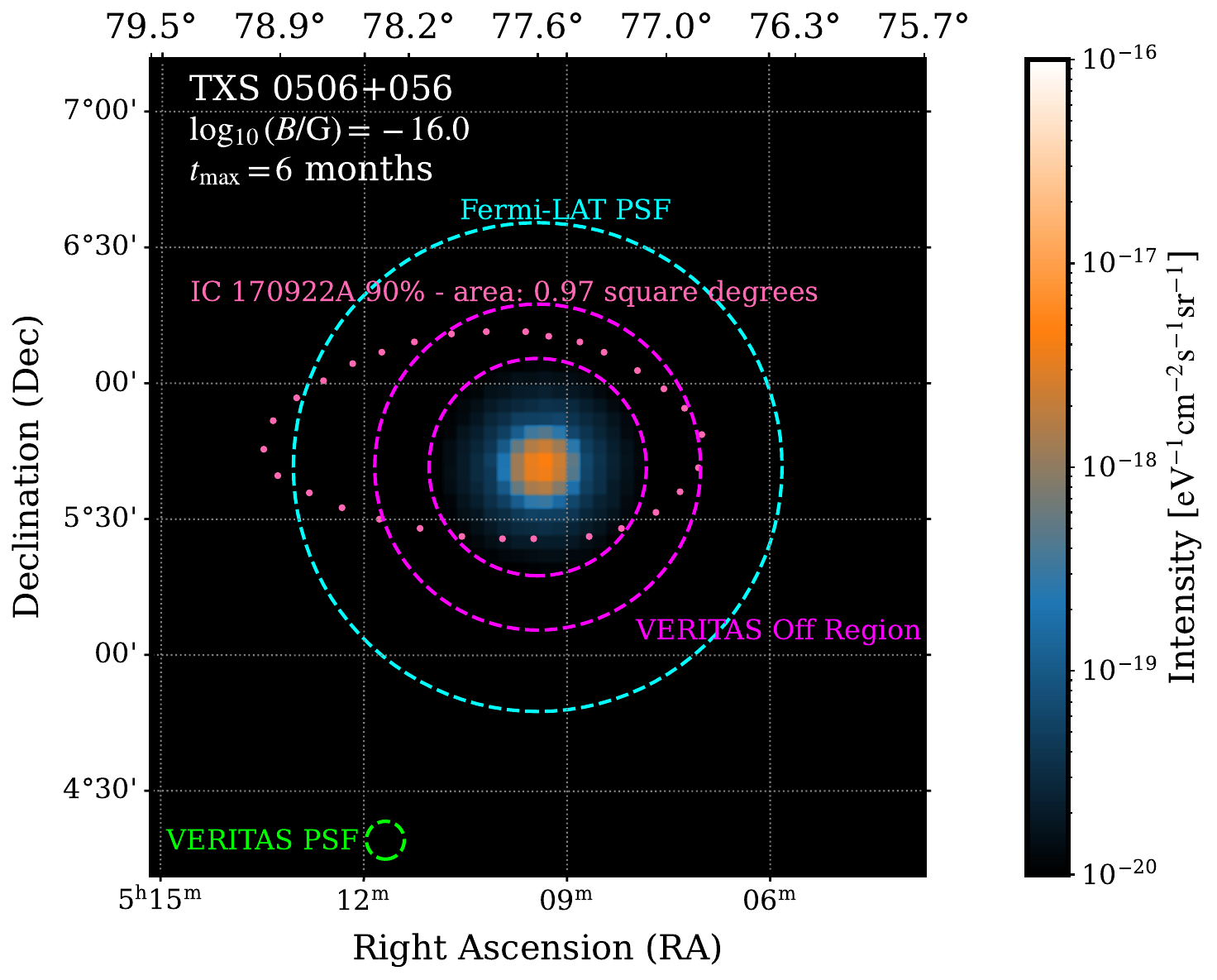}
\includegraphics[width= 1 \linewidth]{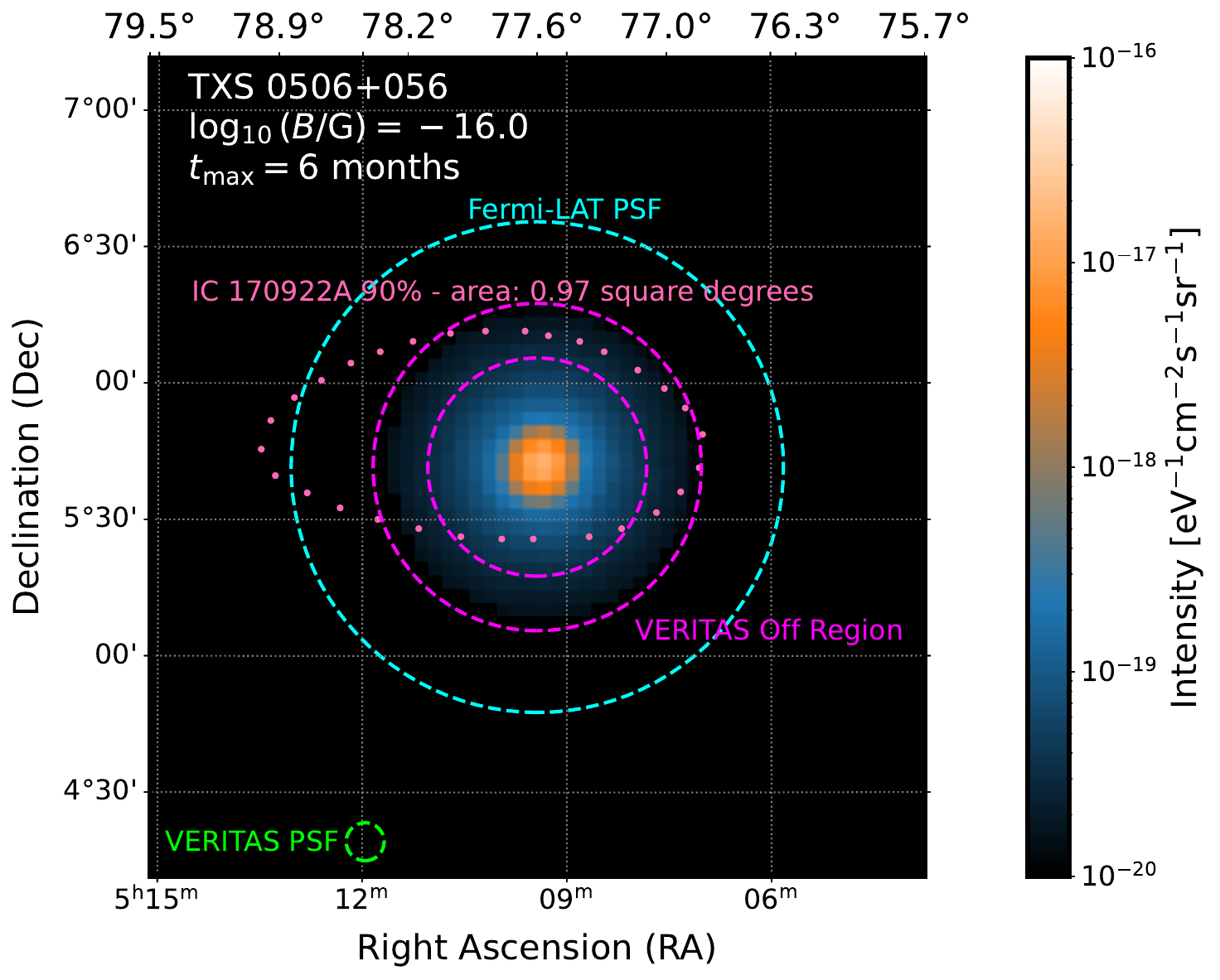}
\caption{The simulated photon cascade flux per solid angle for the best-fit spectrum i.e the subdominant cascade, assuming a LP  $\times$ EBL primary photon  spectrum (top) and PLE  $\times$ EBL primary photon  spectrum (bottom). The values assumed for the maximum blazar activity time and magnetic field are $t_{\text{max}} = 6$~months and $B = 10^{-16} \text{G}$ respectively. The pink dotted line indicates the 90$\%$ confidence-level region for the neutrino event IceCube-170922A (from \cite{2018_Icecube}). The green, cyan, and purple dashed lines represent the PSFs of the VERITAS and the \textit{Fermi}-LAT observations and the VERITAS OFF region, respectively.}
\label{fig:lp_sky}
\end{figure}

\begin{figure}
\centering
\includegraphics[width= 1.1  \linewidth]{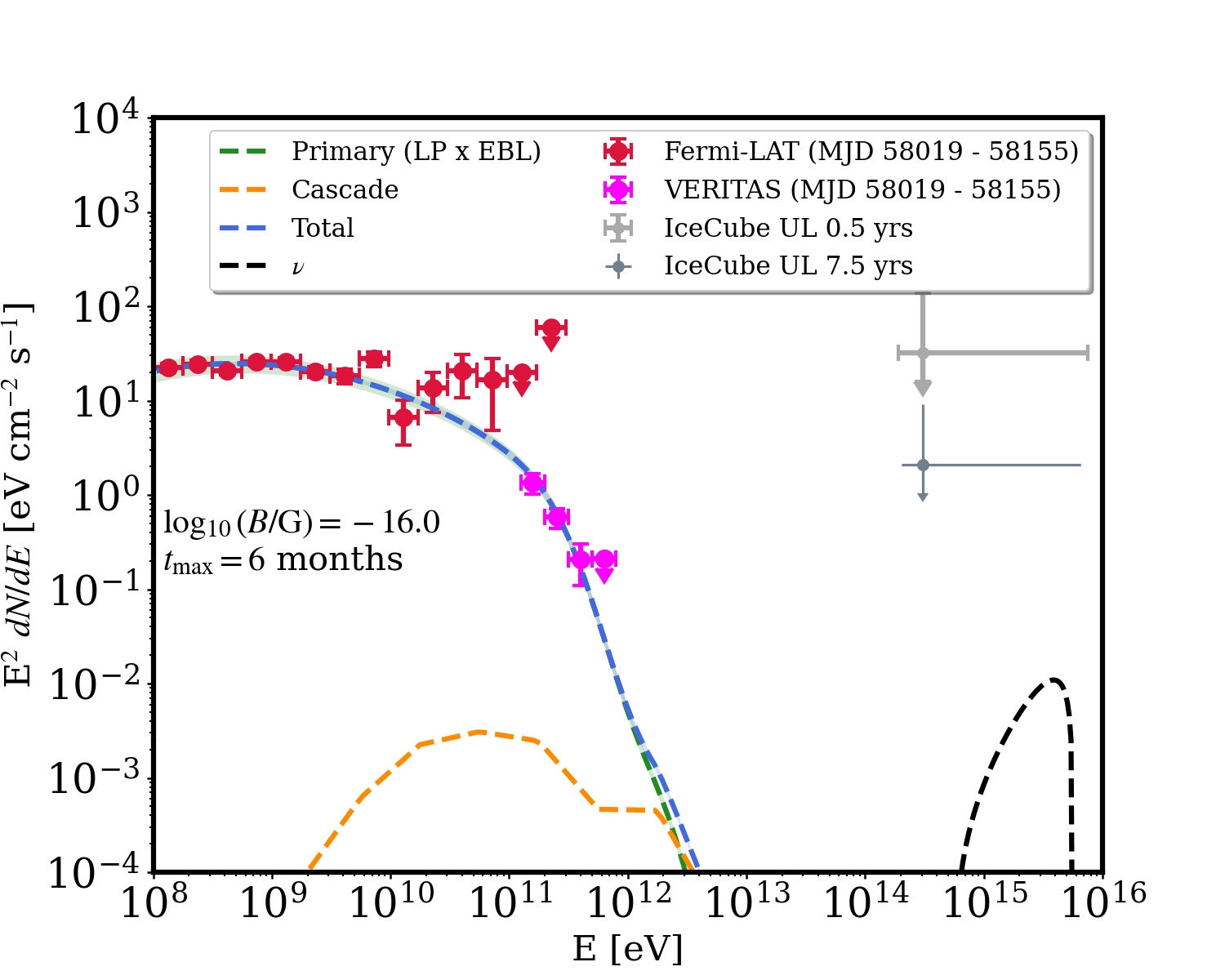}
\includegraphics[width= 1.1 \linewidth]{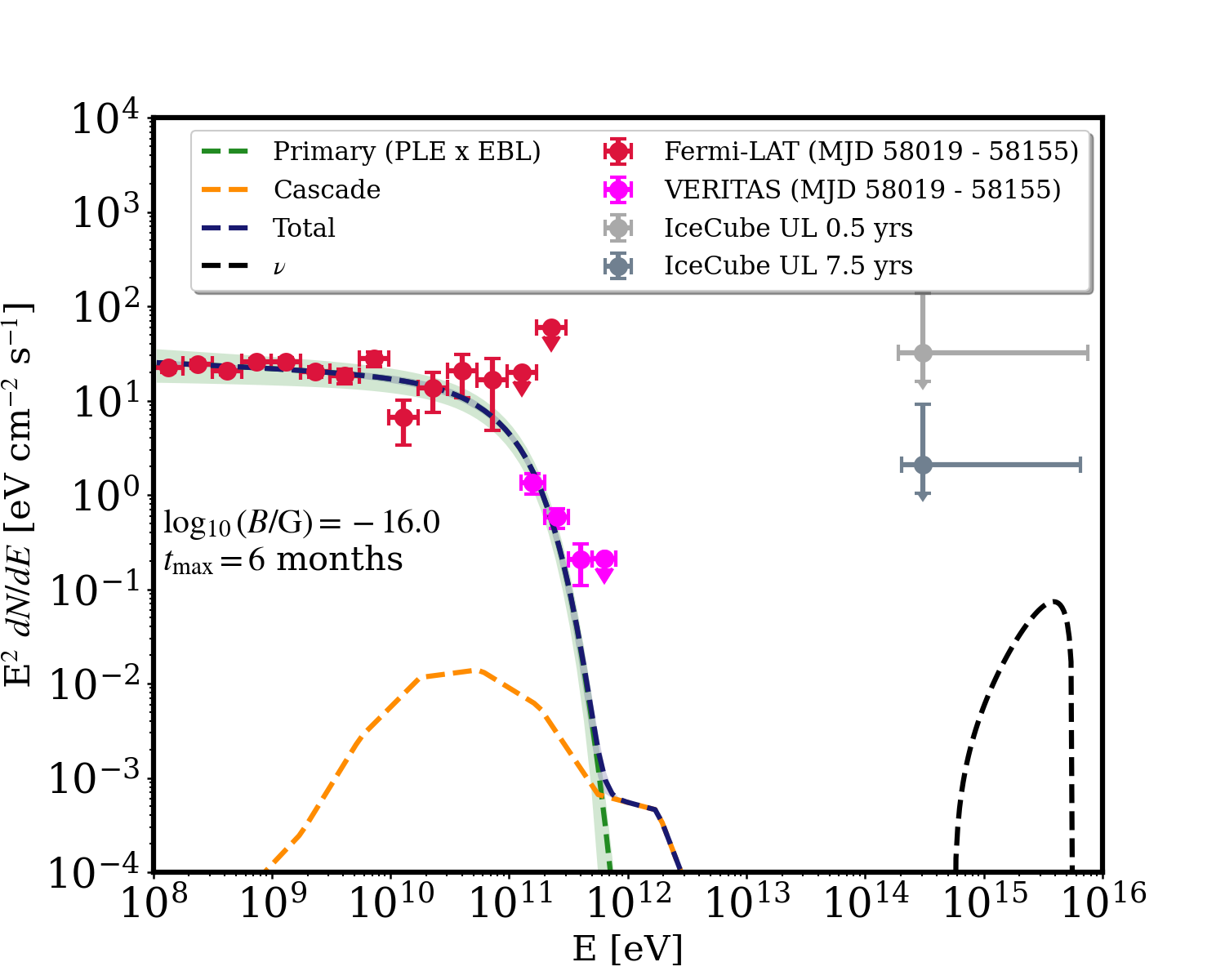}
\caption{The observed \textit{Fermi}-LAT and VERITAS spectra of TXS~0506+056 (red and magenta markers, respectively), together with the total best-fit spectrum (dashed blue line, including uncertainties). This is the sum of the primary  photon spectra LP $\times$ EBL (top) and PLE $\times$ EBL (bottom) (dashed green lines, including uncertainties) and the cascade components (dashed orange lines, including uncertainties) respectively. Spectral points are shown as upper limits if the detection significance is below 2$\sigma$. The SEDs of the predicted cosmogenic neutrinos is shown in black. The spectral curves for the proton cascades and the neutrino spectrum correspond to the best-fit proton injection spectral parameters when considering a LP $\times$ EBL primary photon  spectrum and PLE $\times$ EBL primary photon  spectrum respectively (see Table.~\ref{tab:table}). The values assumed for the maximum blazar activity time and magnetic field are $t_{\text{max}} = 6$ months and $B = 10^{-16} \text{G}$ respectively. Also shown are the representative neutrino flux upper limits that produce on average one detection like IceCube-170922A over a period of 0.5 and 7.5 years assuming an injection spectrum of $dN/dE \propto E^{-2}$ at the most probable neutrino energy, $E_{\nu} = 311$ TeV \citep{2018_Icecube}.}
\label{fig:lp_spectrum}
\end{figure}

We also jointly determine the parameters of the proton injection spectrum and the primary gamma-ray spectrum. The best-fit spectral parameters obtained for the LP primary spectral models, both including and excluding the cascade component, are tabulated in Table~\ref{tab:table}, along with the $\chi_{\text{red}}^{2}$ values obtained for each fit. 
In the case of an LP primary photon spectrum, a combination of proton injection parameters corresponding to the lowest $\chi^{2}$ value are $\alpha_{p} = 2.0$ and $E_{p,\text{max}} = 1.3 \times 10^{16}$~eV. 
The corresponding spectrum, for both the primary and cascade component is shown on the left panel of Fig.~\ref{fig:lp_spectrum}.
The results also show the effect of adding the additional cascade spectral component on the joint fit to the \textit{Fermi}-LAT and VERITAS data. 
The combined gamma-ray spectrum is already well described with the LP $\times$ EBL model ($\chi^{2}/\text{n.d.f}= 1.62$), and the additional cascade component does not improve the fit ($\chi^{2}/\text{n.d.f}= 1.77$). 
The left panel of Fig.~\ref{fig:lp_sky} shows the corresponding simulated photon cascade flux per solid angle for this best-fit spectrum.


\begin{table*}
	\centering
	\caption{
        A combination of best-fit spectral parameters obtained from the joint fit to the combined gamma-ray spectrum for TXS 0506+056, observed with the \textit{Fermi}-LAT and VERITAS. The final column states the  $\chi^{2}/\text{n.d.f}$ values obtained for each model.}
        \hspace{-3 cm}
	\label{tab:table}
	\resizebox{1.1 \linewidth}{!}{
	\centering
	\begin{tabular}{lccccccccccr} 
            \hline
		Fit  &$N_{0}$ &$\Gamma_{\text{LP}}$ &$\beta$ &$E_0$  &$\Gamma_{\text{PLE}}$ &$E^{-1}_{\gamma, \text{cut}}$ &$\alpha_{p}$ &$E_{p, \text{max}}$  & $\kappa$ &$\chi^{2}/\text{n.d.f}$ \\
		function &$[10^{-13} \text{cm}^{-2}\text{s}^{-1} \text{TeV}^{-1}]$ & & &[TeV] &  &[TeV$^{-1}$] & &[eV] & $[\text{cm}^{-2}\text{s}^{-1} \text{eV}^{-1}]$  &\\
		\hline
            LP $\times$ EBL &3.3 $\pm$ 0.5 & 3.14 $\pm$ 0.02 &0.08 $\pm$ 0.05   &1 &- &- &- &- &- &1.62\\
            (LP $\times$ EBL) + Cascade  &3.3 $\pm$ 0.5  & 3.14 $\pm$ 0.02 &0.08 $\pm$ 0.05    &1 &- &- &2.0 &$1.3 \times 10^{16}$ &$6.24 \times 10^{-20}$ &1.77\\
		PLE $\times$ EBL &143.7 $\pm$ 35.4 &- &-  &1 &2.06 $\pm$ 0.03 &12.03 $\pm$ 1.57  &- &- &- &2.53\\
            (PLE $\times$ EBL) + Cascade &147.6 $\pm$ 36.3 &- &-  &1 &2.06 $\pm$ 0.03 &12.74 $\pm$ 1.68 &2.0 &  $7.5 \times 10^{15}$ &$6.24 \times 10^{-19}$ &2.75\\
		\hline
	\end{tabular}}
\end{table*}

\subsubsection{Power law with exponential cut-off primary photon spectrum}

On the bottom panel of Fig.~\ref{fig:lp_contour}, we present the analogous $\Delta \chi^2$-surface obtained assuming a PLE primary gamma-ray spectrum with EBL absorption. In the case of an PLE primary photon spectrum, the parameters of the proton spectrum are more constrained than for the case of a LP primary photon spectrum. Assuming $\delta = 24.2$, the compatible isotropic source-frame proton escape luminosities (corresponding to $\Delta \chi^2 \leq 2.71$ i.e a 90$\%$ confidence level) are between $2 \times 10^{45}$ erg s$^{-1}$ and $3 \times 10^{46}$ erg~s$^{-1}$. 

We again jointly determine the parameters of the proton injection spectrum and the primary gamma-ray spectrum. The best-fit spectral parameters obtained for the PLE primary spectral models, both including and excluding the cascade component, are tabulated in Table~\ref{tab:table}, along with the $\chi_{\text{red}}^{2}$ values obtained for each fit. In the case of an PLE primary photon spectrum, a combination of proton injection parameters corresponding to the lowest $\chi^{2}$ value are $\alpha_{p} = 2.0$ and $E_{p,\text{max}} = 7.5 \times 10^{15}$~eV. 
The corresponding spectrum, for both the primary and cascade component is shown on the right panel of Fig.~\ref{fig:lp_spectrum}.
The results also show the effect of adding the additional cascade spectral component on the joint fit to the \textit{Fermi}--LAT and VERITAS data. 
The combined gamma-ray spectrum is found not to be as well described by the PLE $\times$ EBL model ($\chi^{2}/\text{n.d.f}= 2.53$), as by the LP $\times$ EBL model, and the additional cascade component does not improve the fit ($\chi^{2}/\text{n.d.f}= 2.75$). Nevertheless, we obtain similar exclusion regions for the proton spectrum from both choices of primary model.
The right panel of Fig.~\ref{fig:lp_sky} shows the corresponding simulated photon cascade flux per solid angle for this best-fit spectrum.

We note that, in both cases discussed above, for the proton spectral parameters corresponding to within the $2\sigma$ exclusion region, the contamination of the VERITAS OFF region by cascade photons is estimated to be at most a few percent. This level of contamination does not significantly affect the robustness of our exclusion. However, for parameter values beyond the $2\sigma$ exclusion, the contamination becomes non-negligible, and the corresponding $\Delta\chi^2$ values should be interpreted with caution. A conservative estimate—assuming a VERITAS effective area of $10^9\,\mathrm{cm}^2$ and a total exposure of 35 hours—suggests that the expected number of cascade photons in the OFF region increases from $\sim$0.004 for the best-fit case to $\sim$250 in a scenario near the edge of the $2\sigma$ exclusion, compared to 16,174 total OFF events seen in our analysis.

\subsection{Simulated neutrino spectrum}

For each pair of proton injection parameters $(\alpha_{p}, E_{\text{p,max}})$ used to model the cascade gamma-rays, we also compute the corresponding high-energy neutrino spectrum self-consistently using the \texttt{CRPropa} simulations. That is, the simulated neutrino spectra produced from the same population of protons responsible for the electromagnetic cascades. These spectra depend on both the injection parameters and the ambient photon and magnetic field environments.

In Fig.~\ref{fig:lp_spectrum}, we show the neutrino spectra corresponding to the best-fit proton injection parameters. For context, we also show representative neutrino flux upper limits derived under the assumption of an $E_{\nu}^{-2}$ neutrino spectrum, corresponding to one detection like IceCube-170922A over a period of 0.5 and 7.5 years, respectively \citep{2018_Icecube}. These limits are not part of our model but are included to provide a benchmark for the expected cosmogenic neutrino detectability.
For a proton spectral index of $\alpha_{p} = 2$, the predicted cosmogenic neutrino flux exceeds the IceCube 0.5 year upper limit when the total proton luminosity surpasses $L_{p} \approx 1.4 \times 10^{46} \text{erg s}^{-1}$, thereby placing a further constraint on the allowed proton luminosity in this scenario.

\subsection{Constraints on the fraction of escaped protons}

In the previous Sections, we derived constraints on the spectrum of protons escaping the source, and therefore also provided limits on the luminosity of the escaped protons. We tested both LP $\times$ EBL and PLE $\times$ EBL primary models, each also with an additional cascade component, and obtained the $\chi^2_{red}$ value for each scenario. The constraints were found to depend on the assumed gamma-ray spectrum.
We now change our approach and instead fix the proton spectrum parameters to determine constraints on the fraction of protons that escape into intergalactic space. These protons escape from the acceleration region and contribute towards the observed cosmic-ray spectrum.

\cite{2018_Keivani} model the multiwavelength SED of TXS~0506+056 during the 2017 flare using hybrid leptonic models and the maximum proton injection luminosity for the LMBB2b model was found to be $L_{p}' = 5.4 \times 10^{44}$ erg s$^{-1}$. Considering a Doppler factor, $\delta = 24.2$ \citep{2020ApJ...891..115P}, this corresponds to $L_{p} = 1.9 \times 10^{50}$ erg s$^{-1}$ in the observers' frame. Furthermore, the power-law proton index and the maximum proton Lorentz factor considered in LMBB2b model are $\alpha_{p} = 2$ and $\gamma_{\text{p, max}}' = 1.6 \times 10^{7}$ respectively \citep{2018_Keivani}. In the above, the prime represents quantities being measured in the jet (blob) comoving frame. This maximum Lorentz factor corresponds to a high energy cut-off, $E_{\text{cut}} = \delta \gamma_{p, max}' m c^{2} = 3.6 \times 10^{17}$~eV. For a proton index of  $\alpha_{p} = 2$, this cut-off value is not compatible with our results at a level exceeding 2$\sigma$ (see Fig.~\ref{fig:lp_contour}). 

\begin{figure}
\centering
\includegraphics[width=  \linewidth]{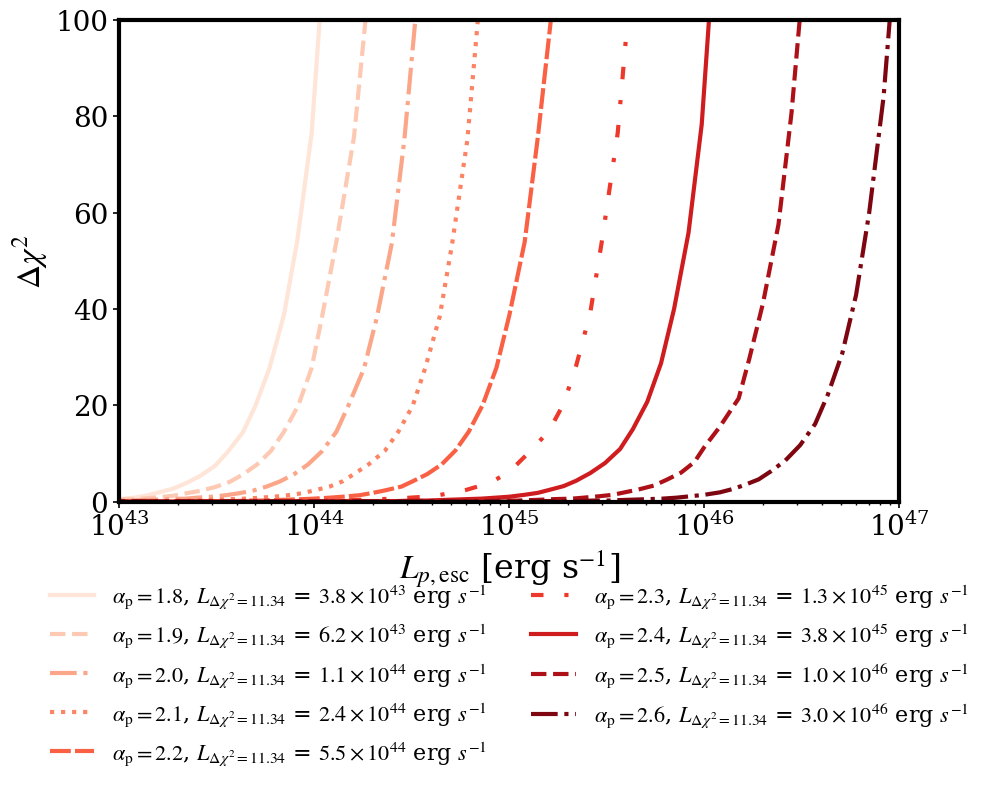}
\caption{The $\Delta \chi^{2}$ values obtained as a function of the proton escape luminosity $L_{p, esc}$. The high-energy cut-off is fixed at $E_{p, \text{max}} = 3.6 \times 10^{17}$~eV and the proton spectral index are fixed in the interval $\alpha_{p} \in [1.8, 2.6]$ at intervals of 0.1. }
\label{fig:lp_contour2}
\end{figure}

In order to constrain the proton escape luminosity, we repeat our fitting procedure considering a (LP~$\times$~EBL)~+~Cascade model, with the high-energy cut-off fixed at $E_{p, \text{max}} = 3.6 \times 10^{17}$~eV. We consider a range of proton spectral indices, for the fixed value of $\alpha_{p} \in [1.8, 2.6]$ at intervals of 0.1. Simultaneously, the cascade emission normalization factor, $\kappa$, is varied, and this allows one to also vary the proton escape luminosity. Fig.~\ref{fig:lp_contour2} shows the $\Delta \chi^{2}$ values obtained as a function of the proton escape luminosity, $L_{\text{p, esc}}$. For a proton spectral parameter $\alpha_{p} = 2$, the value of $L_{p, esc}$ which corresponds to a $\Delta \chi^{2} = 11.34$ (the threshold for a $p$-value $\leq 0.01$ with three free parameters) was found to be $L_{\text{p, esc}} = 1.1 \times 10^{44}$~erg s$^{-1}$. Using the maximum proton injection luminosity  in the observers' frame, $L_{p} = 1.9 \times 10^{50}$~erg s$^{-1}$, this corresponds to a proton escape fraction, $f_{esc} = \frac{L_{\text{p, esc}}}{L_{p}} = 6.1 \times 10^{-7}$. 

This fraction represents a lower limit, as it is derived from a specific maximum proton luminosity model in \cite{2018_Keivani}. For example, \cite{2018_Gao} consider a Doppler factor, $\delta = 26$, maximum proton Lorentz factor, $\gamma_{\text{p, max}}' = 4 \times 10^{5}$ in a hadronic model describing the TXS 0506+056 2017 flare. Applying the same methodology, the maximum proton injection luminosity, $L_{p} =  10^{47}$~erg s$^{-1}$, corresponds to a proton escape fraction, $f_{esc} = 1.1 \times 10^{-4}$. Consequently, while both values indicate that only a small portion of protons injected into the AGN jet escape into intergalactic space under these particular assumptions, the actual escape fraction could be higher depending on the true proton luminosity of the source.

\section{Summary}

In this paper, we report on a new search for proton cascade emission in TXS 0506+056, using a combined dataset from the \textit{Fermi}-LAT and VERITAS. TXS 0506+056 was detected with VERITAS between September 23, 2017 and October 1, 2023 at a significance of 7.5$\sigma$.
We compare the combined gamma-ray spectrum obtained during the six month gamma-ray flare period and the IceCube neutrino upper limits with the predictions of cosmic-ray induced cascades in intergalactic space, obtained using the \texttt{CRPropa} Monte Carlo simulation package.

We perform a joint fit of the primary gamma-ray emission, staying agnostic about its origin, and the proton spectrum. Our method has the advantage that the only assumption we make is that cosmic rays are accelerated in the source and subsequently escape. We apply a statistical analysis to jointly determine the best-fit parameters of a proton emission spectrum describing the data and derive constraints on the proton escape luminosity. The constraints are found to depend on the assumed gamma-ray spectrum with the LP $\times$ EBL primary models providing a better fit than the PLE $\times$ EBL primary models. Moreover, the addition of the cascade components slightly worsens both the LP $\times$ EBL model and PLE $\times$ EBL model fits.

Finally, we fix the proton spectrum parameters to determine constraints on the fraction of protons that escape into intergalactic space. These protons escape from the acceleration region and contribute to the observed cosmic-ray spectrum. We obtain a relatively low proton escape fraction, $f_{esc} =  6.1 \times 10^{-7}$. This fraction represents a lower bound, as it is based on a specific maximum proton luminosity model presented in \cite{2018_Keivani}. While it suggests that only a small fraction of protons injected into the AGN jet escape into intergalactic space under this particular assumption, the actual escape fraction could be higher depending on the true proton luminosity of the source.

Future work towards the study of proton cascade emission will benefit from VHE observations with the upcoming Cherenkov Telescope Array Observatory (CTAO; \cite{2019scta.book.....C}), complemented by next-generation neutrino detectors such as IceCube-Gen2 \citep{2023arXiv230809427I}, KM3NeT (e.g., \cite{2024arXiv240208363A}), and P-ONE \citep{Twagirayezu:2023Sd}. Moreover, a full three-dimensional cascade analysis incorporating realistic background templates and exploring different intergalactic magnetic field distributions will enable even more robust tests of the proton-induced cascade scenario.

\section*{Acknowledgements}
This research is supported by grants from the U.S. Department of Energy Office of Science, the U.S. National Science Foundation and the Smithsonian Institution, by NSERC in Canada, and by the Helmholtz Association in Germany. This research used resources provided by the Open Science Grid, which is supported by the National Science Foundation and the U.S. Department of Energy's Office of Science, and resources of the National Energy Research Scientific Computing Center (NERSC), a U.S. Department of Energy Office of Science User Facility operated under Contract No. DE-AC02-05CH11231. We acknowledge the excellent work of the technical support staff at the Fred Lawrence Whipple Observatory and at the collaborating institutions in the construction and operation of the instrument.

This research was partially supported by NASA grant \emph{NuSTAR} GO-5277. This research has made use of data obtained with \emph{NuSTAR}, a project led by Caltech, funded by NASA and managed by NASA/JPL, and has utilized the \emph{NuSTAR}DAS software package, jointly developed by the ASDC (Italy) and Caltech (USA).

The \textit{Fermi}-LAT Collaboration acknowledges generous ongoing support from a number of agencies and institutes that have supported both the development and the operation of the LAT as well as scientific data analysis. These include the National Aeronautics and Space Administration and the Department of Energy in the United States, the Commissariat à l'Energie Atomique and the Centre National de la Recherche Scientifique / Institut National de Physique Nucléaire et de Physique des Particules in France, the Agenzia Spaziale Italiana and the Istituto Nazionale di Fisica Nucleare in Italy, the Ministry of Education, Culture, Sports, Science and Technology (MEXT), High Energy Accelerator Research Organization (KEK) and Japan Aerospace Exploration Agency (JAXA) in Japan, and the K. A. Wallenberg Foundation, the Swedish Research Council and the Swedish National Space Board in Sweden.

Additional support for science analysis during the operations phase is gratefully acknowledged from the Istituto Nazionale di Astrofisica in Italy and the Centre National d'Etudes Spatiales in France. This work performed in part under DOE Contract DE- AC02-76SF00515.

M.M. acknowledges support from the European Research Council (ERC) under the European Union’s Horizon 2020 research and innovation program Grant agreement No. 948689 (AxionDM).

\section*{Data Availability}

This research has made use of public data and analysis tools provided by the NASA \textit{Fermi} collaboration.
In addition, this work has also made use of the NASA/IPAC Extragalactic Database (NED), which is operated by the Jet Propulsion Laboratory, Caltech, under contact with the National Aeronautics and Space Administration.



\bibliographystyle{mnras}
\bibliography{draft_paper} 

\begin{thebibliography}{}
\makeatletter
\relax
\def\mn@urlcharsother{\let\do\@makeother \do\$\do\&\do\#\do\^\do\_\do\%\do\~}
\def\mn@doi{\begingroup\mn@urlcharsother \@ifnextchar [ {\mn@doi@} {\mn@doi@[]}}
\def\mn@doi@[#1]#2{\def\@tempa{#1}\ifx\@tempa\@empty \href {http://dx.doi.org/#2} {doi:#2}\else \href {http://dx.doi.org/#2} {#1}\fi \endgroup}
\def\mn@eprint#1#2{\mn@eprint@#1:#2::\@nil}
\def\mn@eprint@arXiv#1{\href {http://arxiv.org/abs/#1} {{\tt arXiv:#1}}}
\def\mn@eprint@dblp#1{\href {http://dblp.uni-trier.de/rec/bibtex/#1.xml} {dblp:#1}}
\def\mn@eprint@#1:#2:#3:#4\@nil{\def\@tempa {#1}\def\@tempb {#2}\def\@tempc {#3}\ifx \@tempc \@empty \let \@tempc \@tempb \let \@tempb \@tempa \fi \ifx \@tempb \@empty \def\@tempb {arXiv}\fi \@ifundefined {mn@eprint@\@tempb}{\@tempb:\@tempc}{\expandafter \expandafter \csname mn@eprint@\@tempb\endcsname \expandafter{\@tempc}}}

\bibitem[\protect\citeauthoryear{Aartsen et~al.,}{Aartsen et~al.}{2013}]{PhysRevLett.111.021103}
Aartsen M.~G.,  et~al., 2013, \mn@doi [Phys. Rev. Lett.] {10.1103/PhysRevLett.111.021103}, 111, 021103

\bibitem[\protect\citeauthoryear{Aartsen et~al.,}{Aartsen et~al.}{2014}]{PhysRevLett.113.101101}
Aartsen M.~G.,  et~al., 2014, \mn@doi [Phys. Rev. Lett.] {10.1103/PhysRevLett.113.101101}, 113, 101101

\bibitem[\protect\citeauthoryear{{Aartsen} et~al.,}{{Aartsen} et~al.}{2018}]{2018AdSpR..62.2902A}
{Aartsen} M.~G.,  et~al., 2018, \mn@doi [Advances in Space Research] {10.1016/j.asr.2017.05.030}, \href {https://ui.adsabs.harvard.edu/abs/2018AdSpR..62.2902A} {62, 2902}

\bibitem[\protect\citeauthoryear{{Abeysekara} et~al.,}{{Abeysekara} et~al.}{2018}]{VERITAS_TXS0506_original}
{Abeysekara} A.~U.,  et~al., 2018, \mn@doi [\apjl] {10.3847/2041-8213/aad053}, \href {https://ui.adsabs.harvard.edu/abs/2018ApJ...861L..20A} {861, L20}

\bibitem[\protect\citeauthoryear{{Acharyya} et~al.,}{{Acharyya} et~al.}{2023}]{2023_Qi}
{Acharyya} A.,  et~al., 2023, \mn@doi [\apj] {10.3847/1538-4357/ace327}, \href {https://ui.adsabs.harvard.edu/abs/2023ApJ...954...70A} {954, 70}

\bibitem[\protect\citeauthoryear{{Acharyya} et~al.,}{{Acharyya} et~al.}{2025}]{2025ApJ...982...80A}
{Acharyya} A.,  et~al., 2025, \mn@doi [\apj] {10.3847/1538-4357/adb30c}, \href {https://ui.adsabs.harvard.edu/abs/2025ApJ...982...80A} {982, 80}

\bibitem[\protect\citeauthoryear{{Ackermann} et~al.,}{{Ackermann} et~al.}{2018}]{2018ApJS..237...32A}
{Ackermann} M.,  et~al., 2018, \mn@doi [\apjs] {10.3847/1538-4365/aacdf7}, \href {https://ui.adsabs.harvard.edu/abs/2018ApJS..237...32A} {237, 32}

\bibitem[\protect\citeauthoryear{{Aharonian}}{{Aharonian}}{2002}]{2002_Aharonian}
{Aharonian} F.~A.,  2002, \mn@doi [\mnras] {10.1046/j.1365-8711.2002.05292.x}, \href {https://ui.adsabs.harvard.edu/abs/2002MNRAS.332..215A} {332, 215}

\bibitem[\protect\citeauthoryear{{Aharonian} et~al.,}{{Aharonian} et~al.}{2023}]{2023ApJ...950L..16A}
{Aharonian} F.,  et~al., 2023, \mn@doi [\apjl] {10.3847/2041-8213/acd777}, \href {https://ui.adsabs.harvard.edu/abs/2023ApJ...950L..16A} {950, L16}

\bibitem[\protect\citeauthoryear{{Aiello} et~al.,}{{Aiello} et~al.}{2024}]{2024arXiv240208363A}
{Aiello} S.,  et~al., 2024, \mn@doi [arXiv e-prints] {10.48550/arXiv.2402.08363}, \href {https://ui.adsabs.harvard.edu/abs/2024arXiv240208363A} {p. arXiv:2402.08363}

\bibitem[\protect\citeauthoryear{Aliu et~al.,}{Aliu et~al.}{2014}]{Aliu2014_1ES0229}
Aliu E.,  et~al., 2014, \mn@doi [The Astrophysical Journal] {10.1088/0004-637X/782/1/13}, 782, 13

\bibitem[\protect\citeauthoryear{{Alves Batista} \& {Saveliev}}{{Alves Batista} \& {Saveliev}}{2020}]{2020_Alves_Batista}
{Alves Batista} R.,  {Saveliev} A.,  2020, \mn@doi [\apjl] {10.3847/2041-8213/abb816}, \href {https://ui.adsabs.harvard.edu/abs/2020ApJ...902L..11A} {902, L11}

\bibitem[\protect\citeauthoryear{{Alves Batista}, {Saveliev}, {Sigl}  \& {Vachaspati}}{{Alves Batista} et~al.}{2016a}]{2016PhRvD..94h3005A}
{Alves Batista} R.,  {Saveliev} A.,  {Sigl} G.,   {Vachaspati} T.,  2016a, \mn@doi [\prd] {10.1103/PhysRevD.94.083005}, \href {https://ui.adsabs.harvard.edu/abs/2016PhRvD..94h3005A} {94, 083005}

\bibitem[\protect\citeauthoryear{{Alves Batista} et~al.,}{{Alves Batista} et~al.}{2016b}]{2016JCAP...05..038A}
{Alves Batista} R.,  et~al., 2016b, \mn@doi [\jcap] {10.1088/1475-7516/2016/05/038}, \href {https://ui.adsabs.harvard.edu/abs/2016JCAP...05..038A} {2016, 038}

\bibitem[\protect\citeauthoryear{{Ansoldi} et~al.,}{{Ansoldi} et~al.}{2018}]{ansoldi2018blazar}
{Ansoldi} S.,  et~al., 2018, \mn@doi [\apjl] {10.3847/2041-8213/aad083}, \href {https://ui.adsabs.harvard.edu/abs/2018ApJ...863L..10A} {863, L10}

\bibitem[\protect\citeauthoryear{{Atwood} et~al.,}{{Atwood} et~al.}{2009}]{Fermi_LAT}
{Atwood} W.~B.,  et~al., 2009, \mn@doi [\apj] {10.1088/0004-637X/697/2/1071}, \href {https://ui.adsabs.harvard.edu/abs/2009ApJ...697.1071A} {697, 1071}

\bibitem[\protect\citeauthoryear{{Atwood} et~al.,}{{Atwood} et~al.}{2013}]{atwood2013pass}
{Atwood} W.,  et~al., 2013, arXiv e-prints, \href {https://ui.adsabs.harvard.edu/abs/2013arXiv1303.3514A} {p. arXiv:1303.3514}

\bibitem[\protect\citeauthoryear{{Ballet}, {Bruel}, {Burnett}, {Lott}  \& {The Fermi-LAT collaboration}}{{Ballet} et~al.}{2023}]{2023_dr4}
{Ballet} J.,  {Bruel} P.,  {Burnett} T.~H.,  {Lott} B.,   {The Fermi-LAT collaboration} 2023, \mn@doi [arXiv e-prints] {10.48550/arXiv.2307.12546}, \href {https://ui.adsabs.harvard.edu/abs/2023arXiv230712546B} {p. arXiv:2307.12546}

\bibitem[\protect\citeauthoryear{{Berezinsky}, {Gazizov}  \& {Grigorieva}}{{Berezinsky} et~al.}{2006}]{2006PhRvD..74d3005B}
{Berezinsky} V.,  {Gazizov} A.,   {Grigorieva} S.,  2006, \mn@doi [\prd] {10.1103/PhysRevD.74.043005}, \href {https://ui.adsabs.harvard.edu/abs/2006PhRvD..74d3005B} {74, 043005}

\bibitem[\protect\citeauthoryear{{Berge}, {Funk}  \& {Hinton}}{{Berge} et~al.}{2007}]{2007A&A...466.1219B}
{Berge} D.,  {Funk} S.,   {Hinton} J.,  2007, \mn@doi [\aap] {10.1051/0004-6361:20066674}, \href {http://adsabs.harvard.edu/abs/2007A%26A...466.1219B} {466, 1219}

\bibitem[\protect\citeauthoryear{{Blandford} \& {Levinson}}{{Blandford} \& {Levinson}}{1995}]{Blandford_and_Levinson_1995}
{Blandford} R.~D.,  {Levinson} A.,  1995, \mn@doi [\apj] {10.1086/175338}, \href {https://ui.adsabs.harvard.edu/abs/1995ApJ...441...79B} {441, 79}

\bibitem[\protect\citeauthoryear{{Burrows} et~al.,}{{Burrows} et~al.}{2005}]{2005_Burrows}
{Burrows} D.~N.,  et~al., 2005, \mn@doi [\ssr] {10.1007/s11214-005-5097-2}, \href {https://ui.adsabs.harvard.edu/abs/2005SSRv..120..165B} {120, 165}

\bibitem[\protect\citeauthoryear{Capalbi, Perri, Saija  \& Tamburelli}{Capalbi et~al.}{2005}]{XRT_Reduction}
Capalbi M.,  Perri M.,  Saija B.,   Tamburelli F.,  2005

\bibitem[\protect\citeauthoryear{{Cherenkov Telescope Array Consortium} et~al.,}{{Cherenkov Telescope Array Consortium} et~al.}{2019}]{2019scta.book.....C}
{Cherenkov Telescope Array Consortium} et~al., 2019, {Science with the Cherenkov Telescope Array}, \mn@doi{10.1142/10986.
}

\bibitem[\protect\citeauthoryear{{Cogan}}{{Cogan}}{2008}]{VEGAS}
{Cogan} P.,  2008, in International Cosmic Ray Conference. pp 1385--1388 (\mn@eprint {arXiv} {0709.4233})

\bibitem[\protect\citeauthoryear{{Dom{\'\i}nguez} et~al.,}{{Dom{\'\i}nguez} et~al.}{2011}]{Dominguez_EBL}
{Dom{\'\i}nguez} A.,  et~al., 2011, \mn@doi [\mnras] {10.1111/j.1365-2966.2010.17631.x}, \href {https://ui.adsabs.harvard.edu/abs/2011MNRAS.410.2556D} {410, 2556}

\bibitem[\protect\citeauthoryear{{Donath} et~al.,}{{Donath} et~al.}{2023}]{gammapy:2023}
{Donath} A.,  et~al., 2023, \mn@doi [A&A] {10.1051/0004-6361/202346488}, 678, A157

\bibitem[\protect\citeauthoryear{{Durrer} \& {Neronov}}{{Durrer} \& {Neronov}}{2013}]{2013A&ARv..21...62D}
{Durrer} R.,  {Neronov} A.,  2013, \mn@doi [\aapr] {10.1007/s00159-013-0062-7}, \href {https://ui.adsabs.harvard.edu/abs/2013A&ARv..21...62D} {21, 62}

\bibitem[\protect\citeauthoryear{{Essey} \& {Kusenko}}{{Essey} \& {Kusenko}}{2010}]{2010APh....33...81E}
{Essey} W.,  {Kusenko} A.,  2010, \mn@doi [Astroparticle Physics] {10.1016/j.astropartphys.2009.11.007}, \href {https://ui.adsabs.harvard.edu/abs/2010APh....33...81E} {33, 81}

\bibitem[\protect\citeauthoryear{{Essey}, {Kalashev}, {Kusenko}  \& {Beacom}}{{Essey} et~al.}{2010}]{2010Essey}
{Essey} W.,  {Kalashev} O.~E.,  {Kusenko} A.,   {Beacom} J.~F.,  2010, \mn@doi [\prl] {10.1103/PhysRevLett.104.141102}, \href {https://ui.adsabs.harvard.edu/abs/2010PhRvL.104n1102E} {104, 141102}

\bibitem[\protect\citeauthoryear{Fomin, Fennell, Lamb, Lewis, Punch  \& Weekes}{Fomin et~al.}{1994}]{FOMIN1994151}
Fomin V.,  Fennell S.,  Lamb R.,  Lewis D.,  Punch M.,   Weekes T.,  1994, \mn@doi [Astroparticle Physics] {http://dx.doi.org/10.1016/0927-6505(94)90037-X}, 2, 151

\bibitem[\protect\citeauthoryear{{Gao}, {Fedynitch}, {Winter}  \& {Pohl}}{{Gao} et~al.}{2019}]{2018_Gao}
{Gao} S.,  {Fedynitch} A.,  {Winter} W.,   {Pohl} M.,  2019, \mn@doi [Nature Astronomy] {10.1038/s41550-018-0610-1}, \href {https://ui.adsabs.harvard.edu/abs/2019NatAs...3...88G} {3, 88}

\bibitem[\protect\citeauthoryear{{Garrappa} et~al.,}{{Garrappa} et~al.}{2019}]{2019ApJ...880..103G}
{Garrappa} S.,  et~al., 2019, \mn@doi [\apj] {10.3847/1538-4357/ab2ada}, \href {https://ui.adsabs.harvard.edu/abs/2019ApJ...880..103G} {880, 103}

\bibitem[\protect\citeauthoryear{{Gehrels}}{{Gehrels}}{2004}]{2004AIPC..727..637G}
{Gehrels} N.,  2004, in {Fenimore} E.,  {Galassi} M.,  eds,  American Institute of Physics Conference Series Vol. 727, Gamma-Ray Bursts: 30 Years of Discovery. pp 637--641 (\mn@eprint {} {astro-ph/0405233}), \mn@doi{10.1063/1.1810924}

\bibitem[\protect\citeauthoryear{{Georganopoulos}, {Aharonian}  \& {Kirk}}{{Georganopoulos} et~al.}{2002}]{Georganopoulos_2002}
{Georganopoulos} M.,  {Aharonian} F.~A.,   {Kirk} J.~G.,  2002, \mn@doi [\aap] {10.1051/0004-6361:20020567}, \href {https://ui.adsabs.harvard.edu/abs/2002A&A...388L..25G} {388, L25}

\bibitem[\protect\citeauthoryear{{Halzen} \& {Zas}}{{Halzen} \& {Zas}}{1997}]{1997Halzen}
{Halzen} F.,  {Zas} E.,  1997, \mn@doi [\apj] {10.1086/304741}, \href {https://ui.adsabs.harvard.edu/abs/1997ApJ...488..669H} {488, 669}

\bibitem[\protect\citeauthoryear{{Holder}}{{Holder}}{2011}]{vts_paper}
{Holder} J.,  2011, in International Cosmic Ray Conference. p.~137 (\mn@eprint {arXiv} {1111.1225}), \mn@doi{10.7529/ICRC2011/V12/H11}

\bibitem[\protect\citeauthoryear{{IceCube Collaboration} et~al.,}{{IceCube Collaboration} et~al.}{2018a}]{2018_ICEcube_archive}
{IceCube Collaboration} et~al., 2018a, \mn@doi [Science] {10.1126/science.aat2890}, \href {https://ui.adsabs.harvard.edu/abs/2018Sci...361..147I} {361, 147}

\bibitem[\protect\citeauthoryear{{IceCube Collaboration} et~al.,}{{IceCube Collaboration} et~al.}{2018b}]{2018_Icecube}
{IceCube Collaboration} et~al., 2018b, \mn@doi [Science] {10.1126/science.aat1378}, \href {https://ui.adsabs.harvard.edu/abs/2018Sci...361.1378I} {361, eaat1378}

\bibitem[\protect\citeauthoryear{{IceCube Collaboration} et~al.,}{{IceCube Collaboration} et~al.}{2023}]{2023Sci...380.1338I}
{IceCube Collaboration} et~al., 2023, \mn@doi [Science] {10.1126/science.adc9818}, \href {https://ui.adsabs.harvard.edu/abs/2023Sci...380.1338I} {380, 1338}

\bibitem[\protect\citeauthoryear{{Ishihara}}{{Ishihara}}{2023}]{2023arXiv230809427I}
{Ishihara} A.,  2023, \mn@doi [arXiv e-prints] {10.48550/arXiv.2308.09427}, \href {https://ui.adsabs.harvard.edu/abs/2023arXiv230809427I} {p. arXiv:2308.09427}

\bibitem[\protect\citeauthoryear{{Jorstad} et~al.,}{{Jorstad} et~al.}{2005}]{2005AJ....130.1418J}
{Jorstad} S.~G.,  et~al., 2005, \mn@doi [\aj] {10.1086/444593}, \href {https://ui.adsabs.harvard.edu/abs/2005AJ....130.1418J} {130, 1418}

\bibitem[\protect\citeauthoryear{{Kalashev}, {Kusenko}  \& {Essey}}{{Kalashev} et~al.}{2013}]{2013PhRvL.111d1103K}
{Kalashev} O.~E.,  {Kusenko} A.,   {Essey} W.,  2013, \mn@doi [\prl] {10.1103/PhysRevLett.111.041103}, \href {https://ui.adsabs.harvard.edu/abs/2013PhRvL.111d1103K} {111, 041103}

\bibitem[\protect\citeauthoryear{{Kalberla}, {Burton}, {Hartmann}, {Arnal}, {Bajaja}, {Morras}  \& {P{\"o}ppel}}{{Kalberla} et~al.}{2005}]{column}
{Kalberla} P.~M.~W.,  {Burton} W.~B.,  {Hartmann} D.,  {Arnal} E.~M.,  {Bajaja} E.,  {Morras} R.,   {P{\"o}ppel} W.~G.~L.,  2005, \mn@doi [\aap] {10.1051/0004-6361:20041864}, \href {https://ui.adsabs.harvard.edu/abs/2005A&A...440..775K} {440, 775}

\bibitem[\protect\citeauthoryear{{Keivani} et~al.,}{{Keivani} et~al.}{2018}]{2018_Keivani}
{Keivani} A.,  et~al., 2018, \mn@doi [\apj] {10.3847/1538-4357/aad59a}, \href {https://ui.adsabs.harvard.edu/abs/2018ApJ...864...84K} {864, 84}

\bibitem[\protect\citeauthoryear{{Kopper} \& {Blaufuss}}{{Kopper} \& {Blaufuss}}{2017}]{GCN21916}
{Kopper} C.,  {Blaufuss} E.,  2017, GRB Coordinates Network, 21916, 1

\bibitem[\protect\citeauthoryear{{Krause}, {Pueschel}  \& {Maier}}{{Krause} et~al.}{2017}]{2017APh....89....1K}
{Krause} M.,  {Pueschel} E.,   {Maier} G.,  2017, \mn@doi [Astroparticle Physics] {10.1016/j.astropartphys.2017.01.004}, \href {http://adsabs.harvard.edu/abs/2017APh....89....1K} {89, 1}

\bibitem[\protect\citeauthoryear{Kurahashi, Murase  \& Santander}{Kurahashi et~al.}{2022}]{Kurahashi2022}
Kurahashi N.,  Murase K.,   Santander M.,  2022, \mn@doi [Annual Review of Nuclear and Particle Science] {10.1146/annurev-nucl-011122-061547}, 72, 365

\bibitem[\protect\citeauthoryear{{Li} \& {Ma}}{{Li} \& {Ma}}{1983}]{1983ApJ...272..317L}
{Li} T.-P.,  {Ma} Y.-Q.,  1983, \mn@doi [\apj] {10.1086/161295}, \href {http://adsabs.harvard.edu/abs/1983ApJ...272..317L} {272, 317}

\bibitem[\protect\citeauthoryear{{Maier} \& {Holder}}{{Maier} \& {Holder}}{2017}]{Maier17}
{Maier} G.,  {Holder} J.,  2017, in 35th International Cosmic Ray Conference (ICRC2017). p.~747 (\mn@eprint {arXiv} {1708.04048})

\bibitem[\protect\citeauthoryear{{Mannheim}}{{Mannheim}}{1995}]{1995Mannheim}
{Mannheim} K.,  1995, \mn@doi [Astroparticle Physics] {10.1016/0927-6505(94)00044-4}, \href {https://ui.adsabs.harvard.edu/abs/1995APh.....3..295M} {3, 295}

\bibitem[\protect\citeauthoryear{{Mannheim} \& {Biermann}}{{Mannheim} \& {Biermann}}{1989}]{1989A&A...221..211M}
{Mannheim} K.,  {Biermann} P.~L.,  1989, \aap, \href {https://ui.adsabs.harvard.edu/abs/1989A&A...221..211M} {221, 211}

\bibitem[\protect\citeauthoryear{{Mannheim} \& {Biermann}}{{Mannheim} \& {Biermann}}{1992}]{1992_Mannheim}
{Mannheim} K.,  {Biermann} P.~L.,  1992, \aap, \href {https://ui.adsabs.harvard.edu/abs/1992A&A...253L..21M} {253, L21}

\bibitem[\protect\citeauthoryear{{Padovani}, {Giommi}, {Resconi}, {Glauch}, {Arsioli}, {Sahakyan}  \& {Huber}}{{Padovani} et~al.}{2018}]{2018MNRAS.480..192P}
{Padovani} P.,  {Giommi} P.,  {Resconi} E.,  {Glauch} T.,  {Arsioli} B.,  {Sahakyan} N.,   {Huber} M.,  2018, \mn@doi [\mnras] {10.1093/mnras/sty1852}, \href {https://ui.adsabs.harvard.edu/abs/2018MNRAS.480..192P} {480, 192}

\bibitem[\protect\citeauthoryear{{Paiano}, {Falomo}, {Treves}  \& {Scarpa}}{{Paiano} et~al.}{2018}]{redshift_citation}
{Paiano} S.,  {Falomo} R.,  {Treves} A.,   {Scarpa} R.,  2018, \mn@doi [\apjl] {10.3847/2041-8213/aaad5e}, \href {https://ui.adsabs.harvard.edu/abs/2018ApJ...854L..32P} {854, L32}

\bibitem[\protect\citeauthoryear{{Petropoulou} et~al.,}{{Petropoulou} et~al.}{2020}]{2020ApJ...891..115P}
{Petropoulou} M.,  et~al., 2020, \mn@doi [\apj] {10.3847/1538-4357/ab76d0}, \href {https://ui.adsabs.harvard.edu/abs/2020ApJ...891..115P} {891, 115}

\bibitem[\protect\citeauthoryear{{Poole} et~al.,}{{Poole} et~al.}{2008}]{Poole08}
{Poole} T.~S.,  et~al., 2008, \mn@doi [\mnras] {10.1111/j.1365-2966.2007.12563.x}, \href {http://adsabs.harvard.edu/abs/2008MNRAS.383..627P} {383, 627}

\bibitem[\protect\citeauthoryear{{Protheroe} \& {Szabo}}{{Protheroe} \& {Szabo}}{1992}]{1992PhRvL..69.2885P}
{Protheroe} R.~J.,  {Szabo} A.~P.,  1992, \mn@doi [\prl] {10.1103/PhysRevLett.69.2885}, \href {https://ui.adsabs.harvard.edu/abs/1992PhRvL..69.2885P} {69, 2885}

\bibitem[\protect\citeauthoryear{{Roming} et~al.,}{{Roming} et~al.}{2005}]{2005SS_Roming}
{Roming} P. W.~A.,  et~al., 2005, \mn@doi [\ssr] {10.1007/s11214-005-5095-4}, \href {https://ui.adsabs.harvard.edu/abs/2005SSRv..120...95R} {120, 95}

\bibitem[\protect\citeauthoryear{{Santander}}{{Santander}}{2019}]{santander2019recent}
{Santander} M.,  2019, in 36th International Cosmic Ray Conference (ICRC2019). p.~782 (\mn@eprint {arXiv} {1909.05228}), \url {https://arxiv.org/abs/1909.05228}

\bibitem[\protect\citeauthoryear{{Sch{\"u}ssler} et~al.,}{{Sch{\"u}ssler} et~al.}{2023}]{2023arXiv230915469S}
{Sch{\"u}ssler} F.,  et~al., 2023, \mn@doi [arXiv e-prints] {10.48550/arXiv.2309.15469}, \href {https://ui.adsabs.harvard.edu/abs/2023arXiv230915469S} {p. arXiv:2309.15469}

\bibitem[\protect\citeauthoryear{{Stecker}, {Done}, {Salamon}  \& {Sommers}}{{Stecker} et~al.}{1991}]{1991PhRvL..66.2697S}
{Stecker} F.~W.,  {Done} C.,  {Salamon} M.~H.,   {Sommers} P.,  1991, \mn@doi [\prl] {10.1103/PhysRevLett.66.2697}, \href {https://ui.adsabs.harvard.edu/abs/1991PhRvL..66.2697S} {66, 2697}

\bibitem[\protect\citeauthoryear{{Sunanda}, {Moharana}  \& {Majumdar}}{{Sunanda} et~al.}{2022}]{2022PhRvD.106l3005S}
{Sunanda} {Moharana} R.,   {Majumdar} P.,  2022, \mn@doi [\prd] {10.1103/PhysRevD.106.123005}, \href {https://ui.adsabs.harvard.edu/abs/2022PhRvD.106l3005S} {106, 123005}

\bibitem[\protect\citeauthoryear{{The IceCube Collaboration} et~al.}{{The IceCube Collaboration} et~al.}{2018}]{eaat1378}
{The IceCube Collaboration} et~al., 2018, \mn@doi [Science] {10.1126/science.aat1378}, 361

\bibitem[\protect\citeauthoryear{Twagirayezu, Niederhausen, Sclafani, Whitehorn, Nisa, Yu  \& Halliday}{Twagirayezu et~al.}{2023}]{Twagirayezu:2023Sd}
Twagirayezu J.~P.,  Niederhausen H.,  Sclafani S.,  Whitehorn N.,  Nisa M.,  Yu S.,   Halliday R.,  2023, \mn@doi [PoS] {10.22323/1.444.1175}, ICRC2023

\bibitem[\protect\citeauthoryear{{Wood}, {Caputo}, {Charles}, {Di Mauro}, {Magill}, {Perkins}  \& {Fermi-LAT Collaboration}}{{Wood} et~al.}{2017}]{wood2017fermipy}
{Wood} M.,  {Caputo} R.,  {Charles} E.,  {Di Mauro} M.,  {Magill} J.,  {Perkins} J.~S.,   {Fermi-LAT Collaboration} 2017, in 35th International Cosmic Ray Conference (ICRC2017). p.~824 (\mn@eprint {arXiv} {1707.09551})

\makeatother
\end{thebibliography}

\label{lastpage}
\end{document}